\documentclass{article}

\usepackage{PRIMEarxiv}

\usepackage[utf8]{inputenc} 
\usepackage[T1]{fontenc}    
\usepackage{hyperref}       
\usepackage{url}            
\usepackage{booktabs}       
\usepackage{amsfonts}       
\usepackage{nicefrac}       
\usepackage{microtype}      
\usepackage{lipsum}
\usepackage{fancyhdr}       
\usepackage{graphicx}       
\graphicspath{{media/}}     

\usepackage{scicite}
\usepackage{algorithm}
\usepackage{algorithmicx}
\usepackage{amsmath}
\usepackage{algpseudocode}   
\usepackage{graphicx}
\usepackage{times}
\usepackage{booktabs} 
\usepackage{graphics}
\usepackage{amsmath}
\usepackage{amsfonts}
\usepackage{url} 
\title{\textbf{MiranDa}: \textbf{Mi}micking the Lea\textbf{r}ning Processes of \\Hum\textbf{an} \textbf{D}octors to Achieve C\textbf{a}usal Inference for \\Medication Recommendation
}
\author{
  Ziheng Wang \\
  Division of Biomedical Engineering for Health and Welfare, \\
  Graduate School of Biomedical Engineering \\
  Tohoku University \\
  Sendai, Japan\\
  \texttt{wang@med.tohoku.ac.jp} \\
  \\
  \And
  Xinhe Li \\
  Graduate School of Information Sciences \\
  Tohoku University \\
  Sendai, Japan\\
  \texttt{li.xinhe.c2@tohoku.ac.jp} \\
  \And
  Haruki Momma \\
  Department of Medicine and Science in Sports and Exercise, \\
  Graduate School of Medicine \\
  Tohoku University\\
  Sendai, Japan\\
  \texttt{h-momma@med.tohoku.ac.jp} \\
   \And
  Ryoichi Nagatomi \thanks{\textit{corresponding author}}\\
  Division of Biomedical Engineering for Health and Welfare, \\
  Graduate School of Biomedical Engineering \\
  Tohoku University \\
  \\
  Department of Medicine and Science in Sports and Exercise, \\
  Graduate School of Medicine \\
  Tohoku University\\
  Sendai, Japan\\
  \texttt{nagatomi@med.tohoku.ac.jp} \\
}

\begin{document}
\maketitle

\section*{\centering Highlights}
\begin{minipage}{0.92\linewidth}
\begin{enumerate}
\item[(1)] We provide a paradigm that mimics doctors by utilizing supervised learning and gradient space-based reinforcement learning, which achieves parameter refinements by counterfactual outcomes directly.
\item[(2)] We provided an estimated length of stay in hospital (ELOS) as counterfactual outcomes, differing from the real length of stay in hospital by approximately 0.01 day (1.4 minutes) on average.
\item[(3)] Based on this paradigm, we present a model named \textbf{MiranDa} to offer medication combination recommendations and counterfactual outcomes.
\item[(4)] \textbf{MiranDa} demonstrates significant performance improvements over the Transformer-based baseline. Notably, our model discerns the structural nuances of drug combinations and offers "procedure-specific" attributes.
\vspace{1cm}
\end{enumerate}
\end{minipage}

\section*{\centering Graphical Abstract}
{\centering
\includegraphics[width=0.8\textwidth]{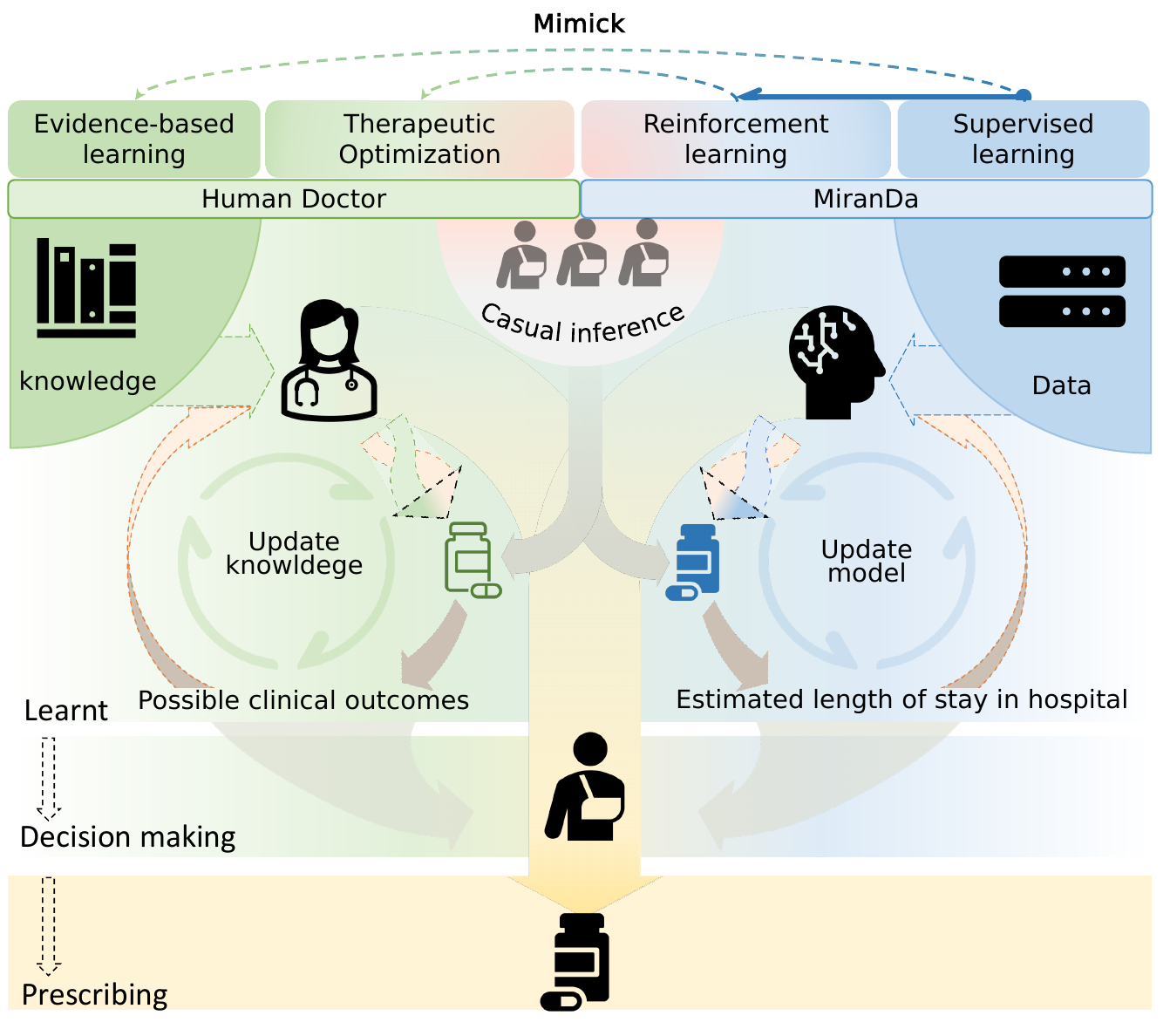}
\par
}
\vspace{1cm}

\begin{abstract}
To enhance therapeutic outcomes from a pharmacological perspective, we propose \textbf{MiranDa}, designed for medication recommendation, which is the first actionable model capable of providing the estimated length of stay in hospitals (ELOS) as counterfactual outcomes that guide clinical practice and model training. In detail, \textbf{MiranDa} emulates the educational trajectory of doctors through two gradient-scaling phases shifted by ELOS: an Evidence-based Training Phase that utilizes supervised learning and a Therapeutic Optimization Phase grounds in reinforcement learning within the gradient space, explores optimal medications by perturbations from ELOS. Evaluation of the Medical Information Mart for Intensive Care III dataset and IV dataset, showcased the superior results of our model across five metrics, particularly in reducing the ELOS. Surprisingly, our model provides structural attributes of medication combinations proved in hyperbolic space and advocated "procedure-specific" medication combinations. These findings posit that \textbf{MiranDa} enhanced medication efficacy. Notably, our paradigm can be applied to nearly all medical tasks and those with information to evaluate predicted outcomes. The source code of the MiranDa model is available at https://github.com/azusakou/MiranDa.
\end{abstract}

\newpage

\section{Introduction}



"Good prescribing" is characterized as maximizing effectiveness, minimizing risks, minimizing costs, and respecting the patient's choices \cite{gooddrug}. Enhancing therapeutic effectiveness and minimizing risk is still a preeminent and difficult concern. Indeed, realizing this goal is presently challenging as medication-related errors persist. For instance, errors occur in 5.6\% of non-intravenous doses and 35\% of intravenous administrations \cite{mcleod2013methodological}. These medication-related errors often result in preventable adverse drug events \cite{ADE1, ADE2, drugrelatedproblem, ADEreason2}, including hospital admissions, extended hospital stays, increased treatment costs, and fatalities \cite{ade3, ade4, ade5}, with mortality rates per 100,000 population ranging from 0.1 to 7.88 \cite{aded1, aded2, aded3, aded4}. The errors frequently correlate with the increasing number of hospitalists practicing immediately after residency, attributed to a deficiency in real-time clinical experience \cite{goodwin2018association}. 
Additionally, the rapid expansion of clinical literature, coupled with increasing specialization, makes it difficult for clinicians to understand medication efficacy comprehensively \cite{bastian2010seventy}. Notably, it is not solely a challenge for neophytes; even seasoned clinicians struggle to determine appropriate prescriptions given lab results, patient vitals, comorbidities, potential drug interactions, and disease progression predictions. Given the complexities involved, achieving optimal prescribing practices is inherently challenging. 

Employed in computer-assisted medical diagnostics, Artificial Intelligence (AI) has emerged as a promising avenue for addressing the challenges of medical aid. Central to this is the Computerized Physician Order Entry, which integrates e-prescribing into Electronic Health Records (EHRs) to bolster prescription safety through drug alerts, detailed medication histories, and the removal of handwritten prescriptions \cite{electronicprescribing}. Recent efforts have harnessed EHRs data to develop efficient artificial intelligence methodologies for medication recommendation \cite{doctorai, micron, medicareai, retain, leap,gamenet, safedrug, csedrug,pccnet, SARMR, grasp, rldrug1,rldrug2,rldrug3}. The principal aim of medication recommendation is to tailor a safe drug regimen for an individual patient, taking into account their specific health conditions. Key milestones include Doctor AI \cite{doctorai} in 2016. Other notable projects include MICRON \cite{micron} which emphasizes enhanced precision, and Medi-Care AI \cite{medicareai}, renowned for bolstering resilience against adversarial training and model interpretability that is also observed in RETAIN \cite{retain}. These approaches aim to provide AI-driven medication recommendations for integration into EHR. 

Recent research has pivoted toward enhanced medication combinations recommendations with drug-drug interaction (DDI) \cite{leap,gamenet, safedrug, csedrug,pccnet, grasp, SARMR}. Several models integrate clinical or molecular knowledge for effective DDI recommendations. LEAP \cite{leap} integrates clinical knowledge into its reinforcement reward to prevent undesirable medication combinations. GAMENet \cite{gamenet} employs a DDI knowledge graph to reduce severe side effects. SafeDrug \cite{safedrug} and CSEDrug \cite{csedrug} both utilize molecular structures to inform their DDI modeling. Specifically, CSEDrug enhances drug encoding and DDI regulation by considering both synergistic and antagonistic DDI, utilizing a graph-centric encoder and multi-faceted loss functions. On the other hand, some models focus on patient-specific data. PCCNet \cite{pccnet} addresses the temporal and spatial changes in medication orders and conditions by considering primary patient medications to mitigate DDIs. GRASP \cite{grasp} leverages patient similarities for representation learning, making predictions based on related patient outcomes and tasks. Moreover, from a representation learning angle, SARMR \cite{SARMR} stands out by generating patient representation distributions and discerning safe combinations using adversarial regularization from raw records. These models integrate various clinical, molecular, and patient-specific data for medication combination recommendations, and improved metric outcomes are consistently achieved.

However, in re-evaluating the notion of "improved metric outcomes", we are faced with two primary ambiguities. 
On one hand, the reliability and efficiency of the medical dataset are unclear. Modeling advancements often lead to improved performance, but overemphasizing better performance may inadvertently foster dependence on imperfect medical datasets, potentially limiting the therapeutic potential of medical interventions. Regrettably, one pivotal yet often overlooked issue is the distinction between natural and medical datasets. Labeling in natural datasets, such as the classification of animals or objects, tends to be straightforward. However, medical datasets, especially those comprising physician diagnoses and treatment details, often require a subjective judgment from medical experts. While label inaccuracies in non-medical datasets range between 0.15\% and 10.12\% \cite{dataseterrors}, ensuring the accuracy and reliability of labels in medical datasets becomes an even more daunting task, given the complexities inherent to medical information. Alarmingly, over 40\% of medical specialists commit prescribing errors, often due to their limited knowledge in the formulation of treatment plans \cite{drugerror1,drugerror2}. Yet, reporting prescribing errors is infrequent, largely due to difficulties in recognition, compounded by fears of litigation and reputational damage \cite{drugreportincident}. Once documented in databases, these errors can misdirect learning models when employing training strategies applicable to natural datasets.

Existing models, despite their ongoing structural innovations, still predominantly rely on feature-to-label mapping. This means that the model primarily learns from the most frequently occurring medication combinations in the outcome data, presenting a foundational challenge in the current datasets. As mentioned earlier, medical datasets are believed to often encompass a range of treatments, from optimal to suboptimal and even incorrect ones. Given this, while a model might avoid some incorrect medication combinations, it may inadvertently downplay the most effective treatment recommendations. Regrettably, the consequences of suboptimal medication combinations go beyond merely prolonging the optimal treatment time. Instances of expensive, prevalent, yet unnecessary medical procedures are not rare \cite{unnessaryoperation}. 

Another reason to re-evaluate the "improved metric outcomes" is the disparity between the effects of prediction based on DDI and the actual expectations of clinicians. One source of disparity is attributable to the oversimplification of DDI. DDI may lead to a spectrum of unexpected side effects, from minor discomforts to severe toxicities. Meanwhile, even with identical side effects, individual reactions to medications can vary widely due to genetic influences \cite{madian2012relating, evans2001pharmacogenomics}. Hence, from a clinical perspective, the mere reduction of DDI to attain what is termed a safer and more efficacious medication presents significant challenges. Another disparity arises during the training of models. It is crucial to underscore that DDI's optimization in predictive models arises during algorithmic training, independent of medications from human doctors.  Overemphasizing DDI could boost certain metrics and DDI rates, but might inadvertently compromise actual medical outcomes by sidelining key considerations like patient preferences and comorbidities. Therefore, we believe a more medically meaningful metric must be proposed to guide the training process.

Regardless of whether the emphasis of the development direction of medication recommendation is on variations in data quality or the specific optimization metrics selected, the terminal concern is the evaluation of the medication-related clinical decision quality. Importantly, the evaluation of quality matches the concept of counterfactual outcomes from causal inference \cite{hernan2019second} which examines how modifying causative factors affects an outcome variable \cite{pearl2009causal}. Existing models focus on several primary aspects: 1) Using encoders to map covariates to representation space, processing combinations, leveraging networks to predict outcomes, and minimizing the distributional distance between factual and counterfactual outcomes \cite{johansson2016learning, fan2021causal}; 2) The theoretical decomposition of covariate relationships \cite{louizos2017causal}; 3) Generating counterfactual output outcomes or achieving balanced representation space distributions  \cite{yoon2018ganite, lorch2022amortized,schwab2018perfect}; 4) Engaging in time series causal learning \cite{lim2018forecasting, rldrug1,rldrug2,rldrug3,sanchez2022causal}. However, in clinical practice, doctors often want to understand counterfactual outcomes: "What will happen if another decision is made?". Despite the emphasis on elucidating causal relationships in most models, the "counterfactual outcomes" for prediction remain obscured. Making these "counterfactual outcomes" transparent could be advantageous, as it would not only directly guide parameter updates but also enable decision modification based on the tangible consequences of those decisions in the real world. Consequently, two pivotal challenges arise: 1) making counterfactual outcomes as a discernible metric and 2) establishing a causal inference-based paradigm that leverages this metric to explore possible outcomes. 

Firstly, this study links counterfactual outcomes and the efficacy of medications. We propose the estimated length of stay (ELOS) as a metric for clinical counterfactual outcomes. In contrast to the length of stay (LOS), an actual medical outcome, ELOS is determined by averaging the LOS of patients exhibiting similar clinical conditions \cite{los_indicator}. Specifically, inspired by approaches in drug repositioning \cite{drugrepositioning} and patient similarity analysis \cite{patientsimilarityanalysis}, we calculate ELOS by identifying analogous patients and medication predictions, ensuring that it dynamically reflects the effectiveness of medication combinations based on real-world medical outcomes. Crucially, integrating this metric within our paradigm ensures seamless compatibility with other modules, such as DDI, present in the architecture.

Next, we investigate achieving a causal inference-based training strategy. This involves two critical aspects: ensuring accurate and reasonable knowledge acquisition, and optimizing this knowledge gleaned through causal inference. Intriguingly, this dual process mirrors the evolution of a physician from a novice to an expert. Typically, a doctor undergoes two main stages of development: the acquisition of evidence-based knowledge in academic settings, followed by therapeutic optimization in clinical environments. The first phase focuses on mastering foundational medical knowledge and techniques. In contrast, the latter delves into diverse clinical outcomes that arise from specific decisions made for individual patients—a process that aligns closely with causal inference. In light of this, we propose a general paradigm for determining the optimal medication combination by mimicking this process. 

While evidence-based training through supervised learning can reliably ensure accurate knowledge acquisition, the real challenge emerges during the therapeutic optimization stage. Notably, we discovered that Reinforcement Learning (RL) \cite{rl98}, a goal-driven learning approach that allows an agent to optimize rewards through interactions with its surroundings, offers the potential to navigate the medical feature space for improved patient outcomes. Indeed, RL might be the simplest way to achieve the causal inference by exploring the different estimated clinical outcomes, as prior work on time series tasks indicates that RL can reduce expected mortality rates \cite{rldrug1,rldrug2,rldrug3}. Thus, we tried to apply RL for therapeutic optimization and explore the gradient space with the guide from ELOS-based reward to make RL suitable for non-time series tasks. However, ELOS is synthesized from patient data and predictions, the retrieved data cannot directly influence gradient updates. Consequently, we use ELOS in a dual capacity: firstly, ELOS-informed perturbations guide the trajectory within the gradient space; secondly, when paired with these perturbations, ELOS serves as a metric for assessing and confirming exploration efficacy, thereby regulating the commencement and conclusion of this exploratory phase. Crucially, the effective assimilation of ELOS into deep learning necessitates the concurrent application of both these functionalities. 

In summary, our objective is to enhance the effectiveness of medication recommendations by addressing dataset constraints and refining optimization metrics. Drawing inspiration from the developmental trajectory of physicians transitioning from novices to experts, we present a causal inference-based paradigm, coupled with a model named MiranDa, which prioritizes the counterfactual outcomes ELOS as the optimization target and integrates supervised learning techniques with reinforcement learning. We hypothesize that 1) ELOS can serve as an effective counterfactual outcome metric for evaluating medication efficacy, and 2) our proposed model will showcase enhanced efficacy in steering medical outcomes by optimizing medication suggestions.

\section{Method}
\subsection{Study design and dataset}
We utilized the Medical Information Mart for Intensive Care (MIMIC) III database \cite{johnson2016mimic, johnson2016mimiciii, goldberger2000physiobank} and IV database \cite{johnson2020mimiciv} to develop and assess our algorithm, primarily aiming to improve recommending medication combinations. The MIMIC-III and MIMIC-IV databases include data for 61,532 and 315,460 unique adult Intensive Care Unit (ICU) admissions, respectively, at Beth Israel Deaconess Medical Center, Boston, MA, USA, spanning 2001–2012 and 2008–2019. We ensured compliance with ethical considerations by securing all necessary permissions for data collection, processing, and dissemination from the appropriate research bodies. All data were stored in a secure environment, facilitating the processes of training, validation, and testing.

To ensure an unbiased evaluation and robust model development, we methodically partitioned the dataset based on individual patients, ensuring that data from each patient was exclusive to a single subset. To validate our concept, we employed the MIMIC-III database, allocating 70\% of the data for training, 15\% for validation, and 15\% for testing, repeated 30 times. For the MIMIC-IV database, we focused more on the trained model by allocating 45\% of the data for training, 5\% for validation, and the remaining 50\% for testing. The development of the algorithm mapping correlations did not involve any data from the testing subsets, which was specifically utilized to evaluate the model performance on unseen data.

\subsection{Data collection and preprocessing}
The admission of each patient to the ICU was assigned a unique ICU stay ID, enabling tracking of distinct ICU episodes. Our inclusion criteria encompassed patients who were aged over 18 years at admission; underwent a continuous regimen of treatment and medication within the study period; and were discharged with an "alive" status. These criteria resulted in the documentation of 35,926 events from 6,154 unique patients in the MIMIC III dataset, and 224,670 events from 119,315 unique patients in the MIMIC IV dataset. Moreover, recognizing the existence of both prevalent and rare cases, we chose not to exclude outliers, even if they could potentially affect performance metrics. Such an inclusive approach offers a more genuine reflection of the patient population.

Data from each hospitalization incorporated a comprehensive history of prior procedures and conditions, as well as patient demographics and clinical characteristics. 
A "LOS" was defined as the period from the onset of a documented hospitalization to the discharge from the ICU. LOS values are normalized to 1.0, which equals 24 hours. The utilization of daily intervals was motivated by the requirement to improve the robustness of the dataset, thus reducing the influence of variables extraneous to clinical decision-making processes
For training our model, we used four input features: diagnoses, procedures, lab events, and patient demographics. The model's prediction target was medication combinations. Medications were coded using the 11-digit version of the National Drug Code. Procedures and diagnoses were categorized using the International Classification of Diseases, Ninth Revision (ICD-9) coding system. Here, "procedures" refer to medical actions undertaken in inpatient hospital settings, and "lab events" denote laboratory test outcomes, such as hematology, chemistry, and microbiology results \cite{johnson2016mimic}. We selected these features due to their clinical relevance and their capability to represent a patient's status accurately. The terminology employed is delineated in Supplementary Terms 1-4.

\subsection{Model design}
\begin{figure*}[!htbp]
	\centering	
    \includegraphics[width=14cm]{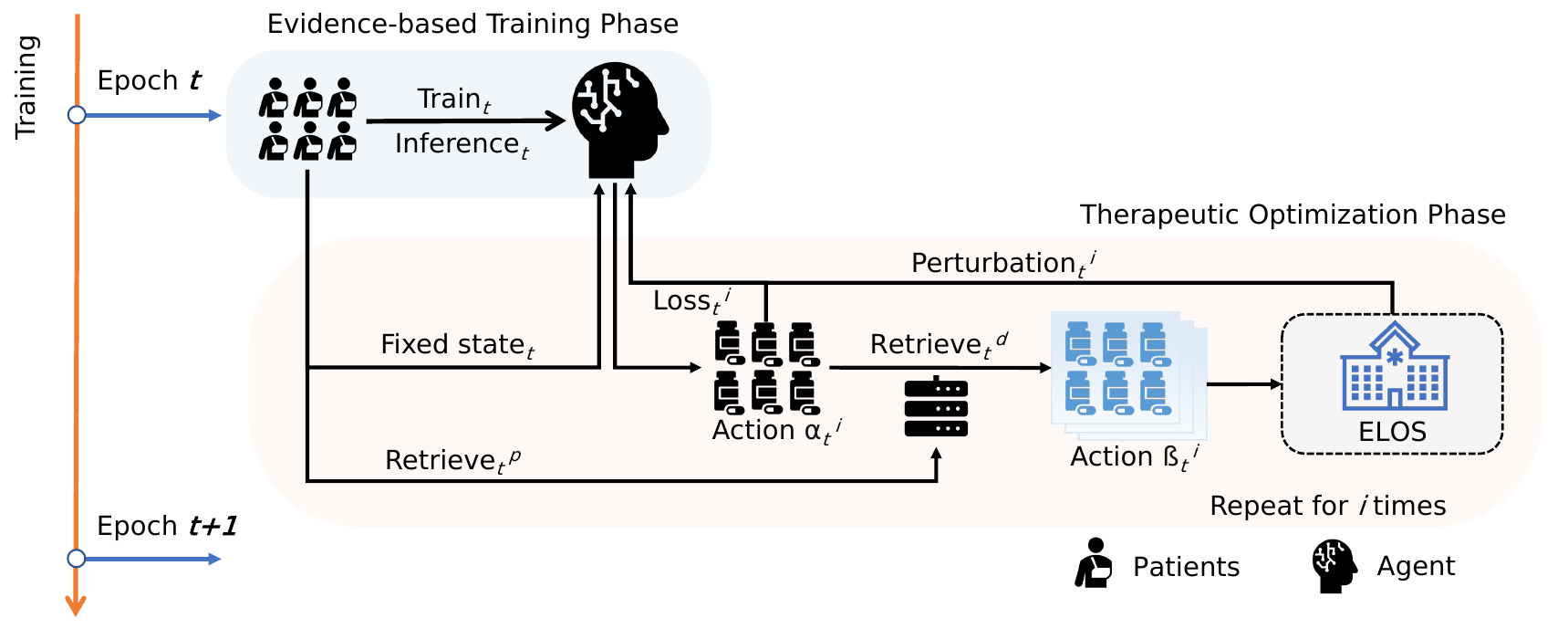}
	  \caption{\textbf{The training process of \textbf{\textbf{MiranDa}}}}
\label{fig:phase2}
\end{figure*}

We introduce \textbf{MiranDa}, a model named for its design principle of \textbf{Mi}micking the lea\textbf{r}ning processes of hum\textbf{an} \textbf{D}octors to achieve c\textbf{a}usal inference for medication recommendation. This model simulates the evolution of clinicians from novices to experts, as shown in Fig. \ref{fig:phase2}. Our approach combines two training phases: the Evidence-based Training Phase (supervised learning) and the Therapeutic Optimization Phase (reinforcement learning in gradient space) for causal inference. The transition between these phases is driven by a clinical outcome-oriented reward ELOS, retrieved from clinical conditions and model predictions. 

\subsubsection{Evidence-based Training Phase}
In the Evidence-based Training Phase, we employ supervised training to establish a mapping relationship between clinical conditions (input features) and medication combinations (predictions). The predictions from this phase serve as an effective starting point, which helps to circumvent the instability introduced by random initialization for the subsequent Therapeutic Optimization Phase.

\begin{figure*}[!htbp]
	\centering	
    \includegraphics[width=16cm]{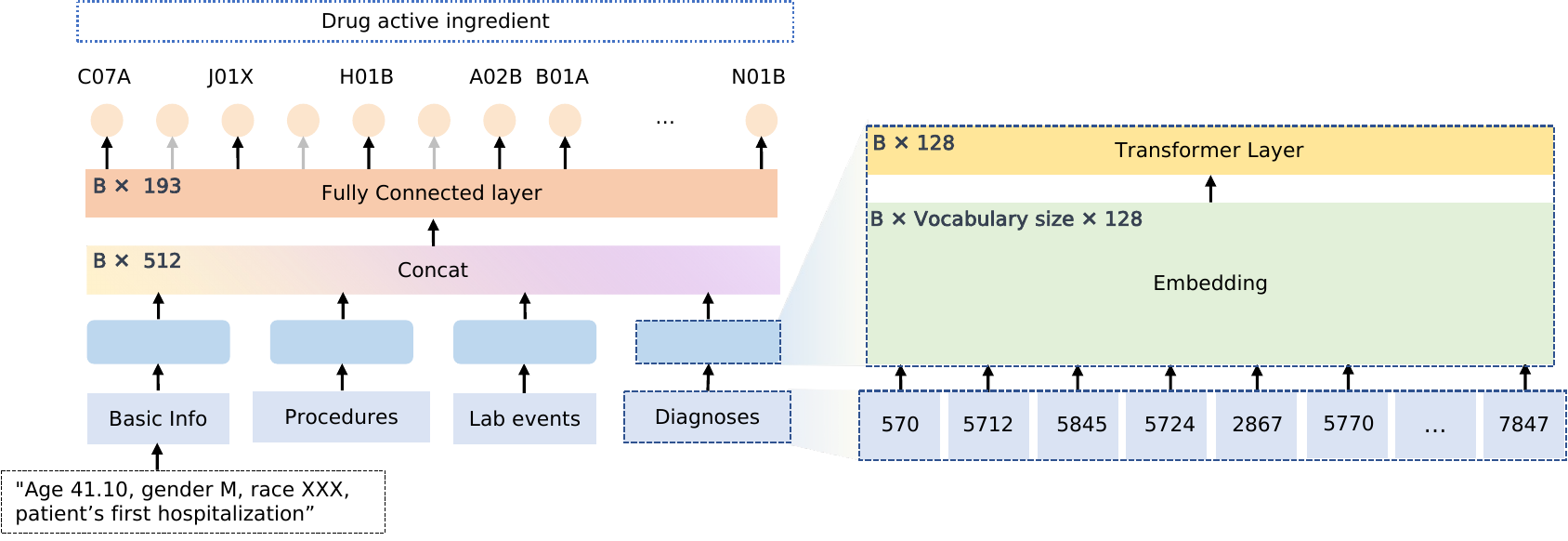}
	  \caption{\textbf{The predictor (agent) of \textbf{MiranDa}}. This predictor is also the baseline of this study.}
\label{fig:phase1}
\end{figure*}

Here, we utilize transformers \cite{attention2017} for the predictor $f\left(\theta\right)$, where \( \theta \) represents the model parameters, the architecture of the predictor is shown in Fig. \ref{fig:phase1}. A patient state \( s_i \), comprising diagnosis, procedures, lab events, and fundamental demographic information which includes age, gender, race, and the patient's sequential number of hospitalizations, the prediction is the medication combination \( \alpha_i \), through \( \alpha_i = f(s_i;\theta) \). The format of inputs is treated as a token. By utilizing the spatial positioning of these individual tokens, the model is facilitated in effectively understanding and processing the associated information. Following the positioning phase, the tokenized features are fed into a transformer layer, generating four distinct vectors representing unique token features in an encoded format. These vectors are concatenated to establish a unified representation, which compactly encapsulates information from all individual vectors. A subsequent fully connected layer reshapes this vector to suit the final output, wherein a softmax function yields a multi-label medication recommendation. Building upon the aforementioned process, the predictor provides a set of medication combinations, denoted as \( \boldsymbol{\alpha} \), where each prediction \( \alpha_i \) belongs to \( \{0, 1\}^{n} \). Here, the value $n$ signifies the total number of medication classes investigated in our study, thereby forming a $n$-dimensional discrete action space.  

\subsubsection{Therapeutic Optimization Phase}

\begin{figure*}[!htbp]
	\centering	
    \includegraphics[width=16cm]{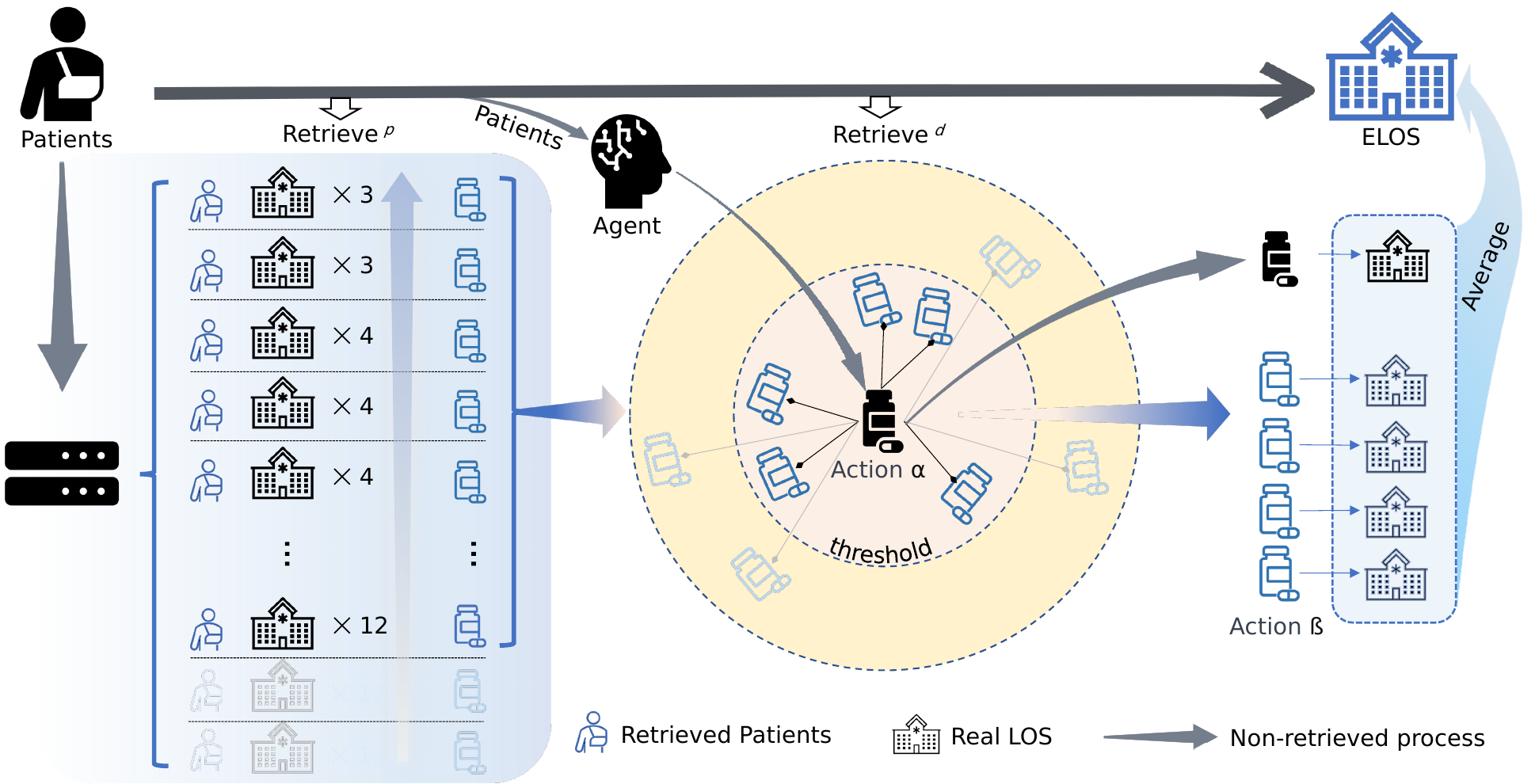}
	  \caption{\textbf{Retrieved process and the calculation of ELOS}}
\label{fig:action}
\end{figure*}

The Therapeutic Optimization Phase involves mimicking the doctor using an agent to explore optimal clinical outcomes. This incorporates the principles of RL for causal inference. In this phase, the predictor $f\left(\theta\right)$ serves as the agent with the patient state $s_i$, diagnosis, procedures, lab events, and patient demographics acting as a fixed state. Actions $\alpha_i$ from the Evidence-based Training Phase are used as initial medication combinations. Based on $\alpha_i$, the auxiliary action space $\beta_i$ is retrieved from similar patients to calculate the reward $\hat{\mathcal{H}}_B$ from ELOS, as shown in fig \ref{fig:action}.
\paragraph{Defining the action space} 

Here, we utilized two retrieve strategies $retrieval^p$ and $retrieval^d$ to get the action $\beta$ based on action $\alpha$. The intrinsic complexities of clinical conditions pose a challenge \cite{intrinsiccomplexities}, as even patients sharing identical diagnoses may exhibit substantial variations in treatments, procedures, and hospitalization duration. To circumvent these intricacies, spurred by concepts such as drug repositioning \cite{drugrepositioning} and patient similarity analysis \cite{patientsimilarityanalysis}, identifying patient trajectories akin to a reference patient aids in predicting the clinical outcomes. In detail, one is $retrieval^p$, which constrains the matched patients with the same procedures and age in a limited range, and the other is $retrieval^d$, which constrains the patients with similar conditions and medications.

Firstly, $retrieval^p$ retrieves similar patients with the same procedures to expand action space $\alpha_i$. In the ICU, on one hand, the patients are particularly vulnerable and frequently require urgent, high-risk therapeutic procedures, which play a crucial and often life-saving role in the management of critical patients \cite{ICUprocedures}. On the other hand, patients undergoing identical procedures often present with analogous health conditions \cite{similarfeatures}. Clinical congruencies are underscored by post-procedure care and monitoring, often meaning that patients face corresponding risks, potential complications, and projected outcomes \cite{careconsensus,postmanagement}. 
Consequently, we use $retrieval^p$ to retrieve the auxiliary action space set $\mathcal{B}$ based on the procedures and utilize a further constraint as the age range within $n$ years of patient $p_i$ with action $\alpha$ from the entire dataset $\mathcal{D}$. We can mathematically express $retrieval^p$ using equation:
\begin{equation}
\mathcal{B}_{p_i} = \left\{ p' \in \mathcal{D} \mid \mathcal{C}(p_i,p') = 1 \right\},\nonumber
\label{retrievalp}
\end{equation}
here, each $p'$ is a patient in $\mathcal{D}$. The function $\mathcal{C}(p_i,p')$ serves as a binary indicator, returning one if patient $p_i$ and patient $p'$ share the exact medical procedures and their ages fall within a specific year difference, and 0 otherwise.

Then, $retrieval^d$ is utilized to constrain the auxiliary action space $\mathcal{B}_{pi}$ drug-wise, further ensuring that the ELOS in the next step are plausible through drug similarity and procedures sameness. This function employs the set of vectors $\mathcal{B}_{pi} = {b_1, b_2, ..., b_n}$, to generate a set $\beta_i$ with a cardinality of $\left|\beta\right| = 50$ that maximizes the cosine similarity between vector $\alpha_i$ and each vector in the set. This can be formalized in the equation:
\begin{equation}
\beta_i = {Top}_\mathcal{K} \left\{ b \in \mathcal{B}_{p_i} \mid \sigma(\alpha_i, b) > \phi, \preceq \right\} , \nonumber
\label{retrievald}
\end{equation}
here the top $\mathcal{K}$ vectors with the highest cosine similarity with $\alpha$ and greater than a threshold $\phi$ are selected. We aim to maximize $\sigma(\alpha_i, b)$ for each $b \in \mathcal{B}$, given $\sigma(\alpha_i, b) > \phi$. A partial order $\preceq$ is defined on $\mathcal{B}$ such that for any $b_i, b_j \in \mathcal{B}$, we have $b_i \preceq b_j$ if and only if $\sigma(\alpha_i, b_i) \leq \sigma(\alpha_i, b_j)$. This detailed mathematical construction provides a clear understanding of the dynamic evolution of the action spaces in our agent.

\paragraph{Counterfactual outcomes and reward}

The reward for each batch of patients, represented by $\hat{\mathcal{H}}_B$, quantifies the efficacy of the actions. Accordingly, $\hat{\mathcal{H}}_B$ computed as:
\begin{equation}
\hat{\mathcal{H}}_B = \frac{1}{N} \sum_{i=1}^{N} \mathcal{R}\left(\alpha_i, \beta_i\right),\nonumber
\label{eq:reward_batch}
\end{equation}
here, this function is based on the mean LOS from each retrieved patient, denoted as $\mathcal{R}$. for each discrete medication combination $\alpha_i$ and its expand drugs set $\beta_i$, we calculate the real LOS, $H\mid\alpha_i$ and counterfactual outcomes as ELOS, $\bar{H}\mid\beta_i^k$, contingent on the action. This difference value is treated as the reward $\mathcal{R}(\alpha_i, \beta_i)$ for the action $\alpha_i$ and $\beta_i$. Mathematically, this can be defined as:
\begin{equation}
\mathcal{R}\left(\alpha_i, \beta_i\right)= H\mid\alpha_i - \bar{H}\mid\beta_i^k,\nonumber
\label{eq:reward_single}
\end{equation}
where $\mathcal{R}\left(\alpha_i, \beta_i\right)$ signifies that shorter hospitalization durations are associated with better medical outcomes. Thus, different from traditional RL, the objective of our agent is to minimize $\hat{\mathcal{H}}_B$, thereby transforming a complex problem of patient recovery into an optimization challenge. This reward-oriented framework provides a concrete and practical target for optimizing our RL agent in the context of pharmaceutical recommendations.

\subsubsection{Training and optimization}

\begin{figure*}[!htbp]
	\centering	
    \includegraphics[width=16cm]{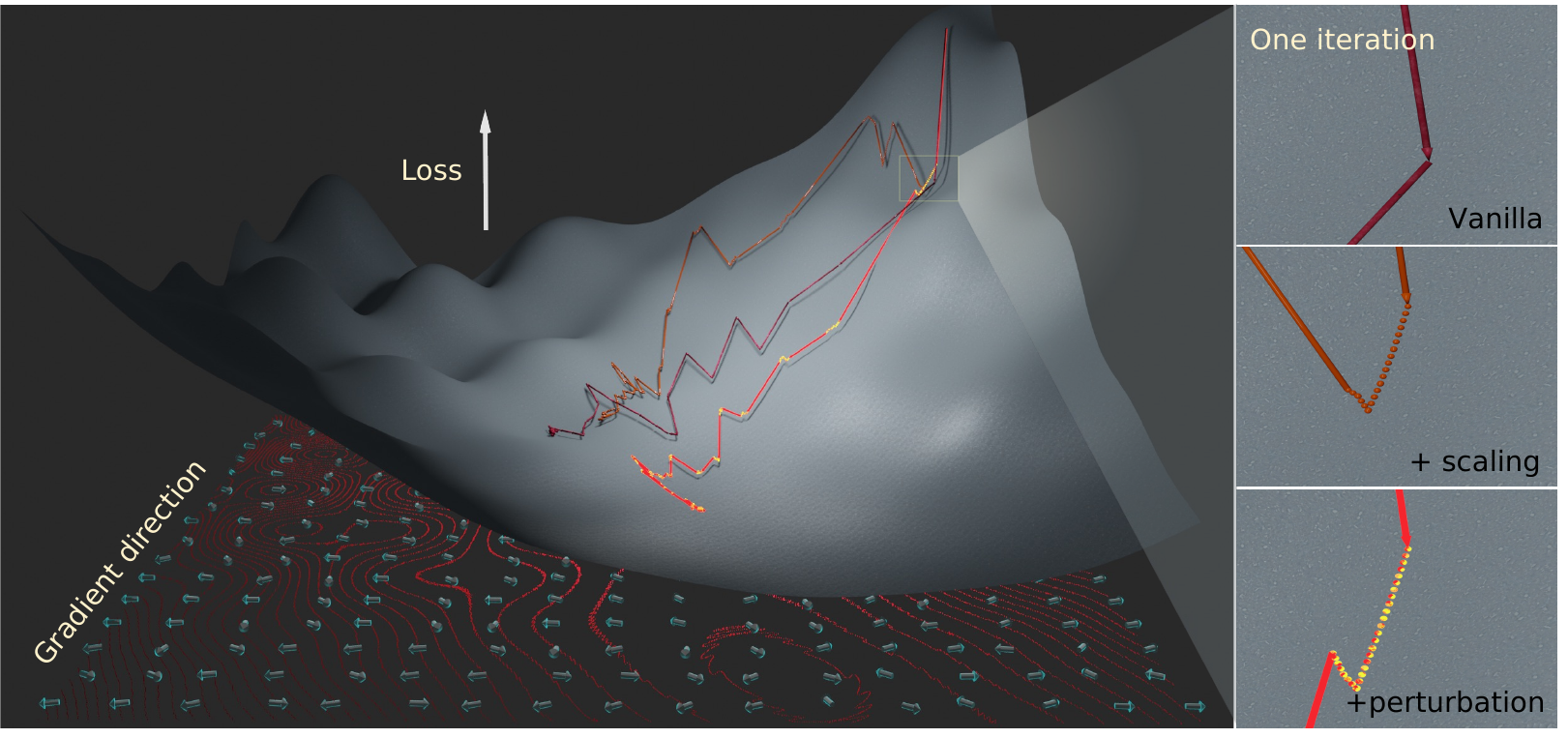}
	  \caption{\textbf{Morphology Comparison among Gradient Descent Strategies: Vanilla, Scaling, and Perturbation}. This figure provides a demonstrative portrayal of gradient descent dynamics employing three illustrative strategies: a straightforward Vanilla approach; Vanilla augmented with Scaling; and an integrated strategy combining Vanilla, Scaling, and Perturbation. For clarity, the right subfigures zoom into selected regions to highlight differences. The red lines and arrows at the bottom of the figure represent the gradient contour plot and the gradient direction of the respective point, respectively.}
\label{fig:grad}
\end{figure*}

We introduce a novel training paradigm that starts from the supervised learning based-start point, we utilize $\hat{\mathcal{H}}_B$ as the indicator to transition to the Therapeutic Optimization Phase and explore the gradient space with the perturbation generated from the reward (rather than the reward itself). The agent employs the minimum ELOS from the retrieved patients – an exploration space reward – as the target, equipping it to explore potential medication combinations under multiple constraints more robustly. The basic concept of exploration in gradient space by the perturbation is shown in Fig. \ref{fig:grad}. 
We proceed with extreme caution when updating parameters, exiting the second stage once a threshold of $\hat{\mathcal{H}}_B$ is reached. This is primarily because models associated with medications should be grounded in existing medical knowledge. In subsequent learning processes, parameters should be cautiously updated, on the premise of ensuring the first-order knowledge can be stably learned or retained \cite{healthcarereinforcement}.

\paragraph{Evidence-based Training Phase loss}

The model parameters first need to be updated with supervised learning. The feed-forward propagation process is represented as a function $f(s; \theta)$, and the error between the model's prediction and the true output is quantified by the binary cross-entropy loss function $l_{BCE}(y, f(X; \theta))$, where $y$ is the true output. Consequently, the Evidence-based Training Phase loss $L(\theta)$ is the average of this binary cross-entropy loss overall training samples. With the reward $\hat{\mathcal{H}}_B$, the loss express as:
\begin{equation}
\mathcal{L}(\theta) = \mathcal{C}(\hat{\mathcal{H}}_B - \delta) \cdot \lambda l_{BCE} + [1 - \mathcal{C}(\hat{\mathcal{H}}_B - \delta)] \cdot l_{BCE},\nonumber
\end{equation}
where $\delta$ is an item that determines whether enters the subsequent phase. The binary function $\mathcal{C}$ is 1 if $\hat{\mathcal{H}}_B \geq \delta$ and 0 otherwise. Consequently, when $\hat{\mathcal{H}}_B \geq \delta$, $L(\theta)$ becomes $\lambda l_{BCE}$, progressing to the Therapeutic Optimization Phase. If $\hat{\mathcal{H}}_B < \delta$, $L(\theta)$ remains $l_{BCE}$, bypassing this phase. A lower value of $\lambda$ thus facilitates more exploratory steps in the Therapeutic Optimization Phase.

\paragraph{Therapeutic Optimization Phase perturbation}
The underlying foundation of our algorithm centers around a tailored Therapeutic Optimization Phase perturbation term, $\mathcal{P}_{RL}(\theta)$, which aims to enhance the exploration space of parameters during updates by applying the derivative of the reward $\hat{\mathcal{H}}_B$.
Here, our main idea was inspired by Stochastic Gradient Langevin Dynamics (SGLD) \cite{sgld}, consequently, our strategy is to replace Gaussian noise with a scaling perturbation term $\mathcal{P}$.
The initial gradient update with perturbation can be expressed as:
\begin{equation}
\theta_{t+1} = \theta_t - \nabla \mathcal{L}(\theta_t) + \nabla \mathcal{P}_{RL},\nonumber
\end{equation}
where $\theta_t$ represents the parameters in the $t$ batch,  $\nabla \mathcal{L}(\theta_t)$ is the gradient of the loss function $\mathcal{L}$ with respect to $\theta_t$. 
The perturbation, $\mathcal{P}_{RL}(\theta)$ is structured with several components designed to promote stability and efficacy in learning. This perturbation incorporates a logarithmic function that engulfs the average reward per batch, weighted by reinforcement confidence, $\gamma$, and subsequently divided by ELOS, denotes by ${\sum_{i=1}^{N}}[\bar{H}|\beta_i^k]$. This logarithmic feature prevents the loss function from increasing indefinitely, thereby constraining the perceived value of escalating rewards. The expression of the perturbation term is as follows:
\begin{equation}
\mathcal{P}_{RL}(\theta) = \log\left(1 + \gamma \left(\frac{ \hat{\mathcal{H}}_B}{{\sum_{i=1}^{N}}[\bar{H}|\beta_i^k]}\right)\right).\nonumber
\end{equation}
Since the update of the perturbation term has a break in the gradient update process for the agent parameters, this better medical outcome is achieved by perturbing the gradient direction. The kernel of the RL perturbation function is $\hat{\mathcal{H}}_B$ divided by ELOS, which aims to maintain stability in the scaling of perturbation medication-wise. This is, essentially, a scaling of the magnitude of the update, i.e., the actual medical significance of $\hat{\mathcal{H}}_B$ as measured by the elos. For example, for the same $\hat{\mathcal{H}}_B$, a smaller ELOS encourages more exploration. Meanwhile, with the reinforcement confidence, $\gamma$, acts as a weighting factor that adjusts the influence of the expected reward on the loss. Moreover, the convex function $\log(x)$ is  for $x > 0$, enabling the application of gradient descent. Therefore, the logarithmic transformation guarantees that the derivative of the perturbation function, $d\mathcal{P}_{RL}(\theta)/d\theta$, remains well-conditioned, fostering a smooth optimization process and precludes abrupt changes in parameter updates during backpropagation.

\paragraph{Perturbed objective function}

Crucially, a perturbed objective function, denoted as $\mathcal{L'}(\theta)$, for updating the parameters in the Therapeutic Optimization Phase, that harmoniously integrates the loss $\mathcal{L}(\theta)$ and the perturbation function $\mathcal{P}_{RL}(\theta)$. The blending of these loss functions is guided by a hyperparameter $\epsilon$, which further ensures the possibility of more steps to explore the optimal medication. This amalgamation is expressed in the form of a convex combination, defined as follows:
\begin{equation}
\mathcal{L'}(\theta) = \epsilon (1-\lambda)\mathcal{L}(\theta) + (1 - \epsilon)\mathcal{P}_{RL}(\theta), \nonumber
\end{equation}
where $\epsilon$ also acts as a balancing factor between the Evidence-based Training Phase loss and the Therapeutic Optimization Phase perturbation, allowing for a smooth transition between the two function types. Crucially, the $\mathcal{L}$ will update for each step in the Therapeutic Optimization Phase perturbation. Our perturbed objective function, $\mathcal{L'}(\theta)$, thus encapsulates the strengths of both conventional mapping relations (represented by the standard loss) and the space that may have optimal medical outcomes (captured by the RL perturbation). This balance is crucial for the agent deployed in dynamic environments such as ICU settings, where actions will not harm the parameters even if they diverge from general mapping relations. Since even if the ELOS is not better, the parameters can still be updated to the directions with a lower loss, and the "mission" of better medical outcomes could be achieved in the further batch. 
For a comprehensive step-by-step procedure of our \textbf{MiranDa} approach, refer to Algorithm~\ref{alg:myalgorithm}.
\begin{algorithm}
\caption{The \textbf{MiranDa}}\label{alg:myalgorithm}
\begin{algorithmic}[1]
\Statex \textbf{Input:} State variables, \( s \)
\Statex \textbf{Initialize:} Parameters, \( \theta \)
\For{each epoch}
    \For{each batch}
    
        \Comment{Evidence-based Training Phase starts}
        \State Predict action (predictions), \( \alpha = f(s;\theta) \)
        \State Calculate loss, \( L(\theta) \)
        \For{each patient in batch}
            \State  Generate \( \beta_i \) from  \( \alpha_i \) by retrieve
        \EndFor
        \State Calculate reward \( \hat{\mathcal{H}}_B \) 
        
        \Comment{Therapeutic Optimization Phase starts}
        \If{ \( \hat{\mathcal{H}}_B \) exceeds \( \delta \)}
            \State  Update \( \theta \gets \theta - \eta \nabla \lambda \mathcal{L}(\theta) \)
            \Repeat
                \State Generate perturbation \( \mathcal{P}_{RL}(\theta) \) 
                \State Calculate objective function \( \mathcal{L'}(\theta) \)
                \State Update \( \theta \gets \theta - \eta \nabla \mathcal{L'}(\theta) \)
                \State Update \( \alpha, \beta, \hat{\mathcal{H}}_B \)
                \State Update \( \mathcal{L}(\theta) \)
            \Until{ \( \hat{\mathcal{H}}_B \) becomes negative}
        \Else
            \State Update \( \theta \gets \theta - \eta \nabla \mathcal{L}(\theta) \)
        \EndIf
    \EndFor
    \If{stopping criteria are met}
        \State Exit loop
    \EndIf
\EndFor
\Statex \textbf{Output:} action (predictions), \( \alpha_i \); counterfactual outcomes, \(\bar{H}|\beta_i^k \)
\end{algorithmic}
\end{algorithm}

\subsubsection{Evaluation metrics}
We evaluate our model using a diverse set of metrics to ensure a comprehensive and robust assessment. The ELOS serves as a critical measure of the model's capability in suggesting medication combinations. It indicates not just the success of a patient's recovery, but also the counterfactual outcomes. ELOS thus directly measures model performance with the clinical aim of minimizing ICU durations to improve patient well-being. We further incorporate the Receiver Operating Characteristic Area Under Curve (ROC AUC) to understand the model's accuracy across different thresholds \cite{AUC}. The Precision-Recall AUC (PR AUC) evaluates the precision-recall trade-off, which is especially important for imbalanced datasets \cite{PRAUC}. Together, ROC AUC and PR AUC offer a comprehensive perspective on predictive accuracy. The F1 score balances precision and recall \cite{F1scores}, and we will also report separate Precision and Recall scores for deeper insight \cite{precisionrecall}. The Jaccard Index measures the similarity between predicted and actual medication combinations, indicating the model's recommendation accuracy \cite{Jaccard}. Additionally, we monitor the DDI rate \cite{SSI-DDI}. Through these metrics, we provide a thorough evaluation of the model, showcasing its efficacy and safety.

\subsection{Data interpretation}
We interpreted the intrinsic correlations among various medication combinations. These combinations were derived from three primary sources: our predictive model, baseline and human doctors. We aimed to elucidate the relationships between these medication combinations and several other variables, including patient diagnosis, procedures, and demographic information. We achieved this using a three-fold methodology: structuring our data in hyperbolic space, clustering the data into discernible categories, and interpreting the clusters.

\subsubsection{Spatial positioning in Hyperbolic Geometry}

Hyperbolic Geometry, a non-Euclidean geometry that excels at capturing and representing data with inherent hierarchical structures \cite{hyperbolic,hyperbolicmapping}, offers a unique lens to discern hierarchies in medication combinations. Transitioning high-dimensional data to this geometry preserves the inherent relationships and distances, a challenge often encountered in traditional Euclidean representations \cite{dimensionality, nonlinear}. Therefore, we utilized the Hyperbolic model and its representation in the Poincaré Disk Model in this study, where each point symbolizes a unique medication combination. The spatial positioning in this space reflects the similarities and disparities of the underlying policies \cite{hyperbolicspace}, instrumental for analyses that capture the genuine relationships within the dataset.

\paragraph{Spatial positioning in Hyperboloid Model}

To achieve spatial positioning in hyperbolic space, we employed the Uniform Manifold Approximation and Projection (UMAP) algorithm, renowned for its effectiveness in capturing hierarchical and topological data properties \cite{umap}. UMAP constructs a fuzzy simplicial set representation to approximate the manifold structure of $\alpha$. By identifying neighbors of each point in $\alpha$ within the high-dimensional space, a weighted graph, $\mathcal{G}$, is constructed. The weights of $\mathcal{G}$ indicate the likelihood of data points being neighbors and are described by:
\begin{equation}
\mathcal{G} = { (\alpha_i, \alpha_j) | \alpha_i, \alpha_j \in \alpha \land d(\alpha_i, \alpha_j) < \epsilon }, \nonumber
\end{equation}
here, $d(\alpha_i, \alpha_j)$ represents a distance metric attentive to local structures, while $\epsilon$ is a threshold for proximity. The objective is to reduce the cross-entropy between $\mathcal{G}$ and its hyperbolic counterpart, ensuring the preservation of both local and global structures. Given the spatial positioning in $\mathbb{H}^3$ as $\mathcal{HS}$, the optimization is:
\begin{equation}
\mathcal{HS}^* = \text{argmin}_{\mathcal{HS}} CE(\mathcal{G}, \mathcal{HS}), \nonumber
\end{equation}
where $\mathcal{HS}^*$ represents the optimal hyperbolic spatial positioning that minimizes the divergence $CE$ between $\mathcal{G}$ and $\mathcal{HS}$.

\paragraph{Transition from Hyperboloid Model to Poincaré Disk Model}
To transition data from the Hyperboloid Model to the Poincaré Disk Model, we employed stereographic projection \cite{Poincaré}. The Poincaré Disk Model represents hyperbolic geometry using points within a disk, with geodesics portrayed as circle segments orthogonal to the disk boundary or as its diameters \cite{poincareclustered}. This representation is advantageous for visualizing intricate datasets in 2D while retaining data nuances.

A point, $h_i$, in the hyperboloid model is mapped to its counterpart, $p_i$, in the Poincaré disk model. The transformation is expressed by $p_i = \frac{h_i}{1 + h_3}$, where $i = 1,2$. Each coordinate of $h_i$ in $\mathbb{H}^3$ is divided by the sum of one and its third coordinate. After this projection, data is visualized as a scatter plot within the boundary of the Poincaré disk, which acts as the line at infinity. In this representation, the proximity between data points underscores their similarity, revealing the manifold structure of the original data.

\subsubsection{Data segmentation}
For data segmentation, we utilized the K-means clustering algorithm \cite{kmeans1982} on the hyperbolic embeddings of medications. This algorithm partitions data into "K" non-overlapping clusters based on inherent data similarities. Applying K-means to the Poincaré Disk Model ensures an accurate representation of the true data groupings, benefiting from the preserved hierarchical relationships. This method enhances the granularity of subsequent theme extraction.

To determine the optimal number of clusters, we calculated the Bayesian Information Criterion (BIC) for potential cluster counts ranging from 10 to 30. Based on the BIC values, we selected ten as the optimal number of clusters, guided by the inherent patterns within our dataset. After forming the clusters, we represented our data in both the hyperboloid and Poincaré disk models. For clearer visualization, each model assigns a unique color to each cluster, providing structure to our data representation.

\subsubsection{Theme extraction}
After segmentation, we employed the Term Frequency-Inverse Document Frequency (TF-IDF) value \cite{tf_idf_ref} to ascertain the significance of words (e.g., medications or procedures) in a document relative to a corpus. The TF-IDF values of all words in each cluster delineated the themes of medication, diagnostic, procedures, and lab events.

Given a document $d$ from the document set $D$ and a term $t$ from the term set $T$, the TF-IDF is computed as the product of two metrics: term frequency (TF) and inverse document frequency (IDF). The TF metric represents the frequency of a word within a particular document, while the IDF metric measures the word's informative value by evaluating its rarity across the entire document set. The TF-IDF formula is:
\begin{equation}
\begin{split}
TF-IDF(t, d, D) = \left( \frac{f_{t, d}}{\max { f_{t', d} : t' \in d }} \right) \times \left( \log \left( \frac{|D|}{| { d \in D : t \in d } | + 1} \right) + 1 \right),
\end{split}
\notag
\end{equation}
here, the term frequency $f_{t, d}$ of term $t$ in document $d$ is normalized by the peak frequency of any term in $d$. The normalized TF is then multiplied by the IDF, calculated as the logarithm (base 10) of the total document count $|D|$ divided by the count of documents containing term $t$, $|{d \in D : t \in d}|$.

\subsubsection{Hierarchical thematic representation}
We employed a Hierarchical Thematic Representation with a dendrogram to display clusters based on dominant themes. This method confirms not only the structural attributes of medication combinations but also illuminates hierarchical inter-cluster relationships.

Using the TF-IDF measure from a diverse document corpus, we established term importance within labels. From this matrix, we computed a linkage matrix using the complete linkage method, which considers the maximum distances between observation pairs. This linkage matrix facilitates the creation of a hierarchical clustering dendrogram. Each leaf node represents a theme word, while internal nodes denote theme word clusters. The height of a node signifies the distance at which its cluster formed, indicating the similarity or divergence between theme words.

\subsubsection{Co-occurrence}
We visualized the relationships of medications and procedures using a technique referred to as the co-occurrence plot. We define "co-occurrence" such that for each unique hospitalization of a patient, we examine all procedures and medications associated with that hospitalization. If procedure "x" and medication "y" appear together during this hospitalization, the corresponding value in the co-occurrence matrix increments by one. We computed co-occurrences for both drug-drug and drug-procedure pairs using the complete result sets from the \textbf{MiranDa}, baseline, and human doctor. This graphical representation allows us to identify patterns, relationships, and dependencies between various drugs and their associated procedures, as well as among different medications. 

\subsection{Statistical analysis} 
For a rigorous evaluation of the differences between the baseline model and the \textbf{MiranDa} model, we utilized a statistical analysis strategy grounded in repeated random sampling. We randomly partitioned the data 30 times; therefore, the effective sample size ($n$) for our analyses was 30.

To ascertain if there was a statistically significant difference in performance between the two models, we applied the paired \textit{t}-test. This test was selected for its capability to compare means from paired data or repeated measurements on the same subjects. In our context, it allowed for comparing the baseline and \textbf{MiranDa} models using identical data splits, thereby controlling for potential sources of variability inherent to the data. A \textit{p}-value ($p<0.05$) from the paired \textit{t}-test would suggest a significant performance discrepancy between the two models.

\subsection{Implementation details}
The experiments were executed on a computational framework comprising a single GeForce GTX 3080 GPU workstation with a 10 GB memory provision. The models underwent a training period of 50 epochs, with an early stopping mechanism initiated from the fifth epoch, set to withstand a tolerance of three epochs. Decay factor $\lambda$, threshold $\delta$, Confidence $\gamma$, Blend factor $\epsilon$ was 0.9, 0.2, 0.5, 0.2. The chosen optimization algorithm was Adam, operating at a learning rate of $1 \times 10^{-3}$. A batch size of 512 was uniformly adhered to across the experiment. 

The data stratification approach relied on patient-specific segregation as opposed to hospitalization-based segmentation. We refrained from using cross-validation in our performance assessment to mitigate bias from overlapping samples. Instead, we utilized a range of random seed values from 0 to 29, facilitating 30 distinct dataset splits into training, validation, and test sets for MIMIC III database. This approach ensured a more robust evaluation of the model's performance by allowing repeated and independent data splitting. The results from the random seed 0 were used for visualizing and explaining the model. Meanwhile, random seed 0 was utilized to split MIMIC IV database.

The experiments were programmed using Python 3.9.0 for its extensive support and compatibility with scientific computation libraries. Deep learning applications were carried out with PyTorch, renowned for its dynamic computational graph and efficient memory usage. The development was largely facilitated by pyHealth \cite{pyhealth2023yang}, a Python-based open-source healthcare data analytics library, which provides the foundation for structuring and modeling. 
The 3D visualizations were rendered using Blender (Version 3.5.1).

\section{Results}
The comprehensive dataset from MIMIC-III and MIMIC-IV, included a variety of parameters such as diagnosis, procedures, medications, lab events, LOS, gender, ethnicity and age, as summarised in Table \ref{tab:mimic-iii}.

\begin{table*}[!htbp]
  \centering
  \caption{\textbf{Summary of datasets statistics}}
  \resizebox{\textwidth}{!}{
    \begin{tabular}{ccccccccccc}
    \toprule
    
     & \multicolumn{8}{c}{MIMIC-III Dataset (Events: 35,926, Patients: 28,345)} \\
     \cmidrule(lr){2-9} 
     \multicolumn{1}{c}{} & \multicolumn{1}{c}{Diagnosis} & \multicolumn{1}{c}{Procedures} & \multicolumn{1}{c}{Medications} & \multicolumn{1}{c}{Lab events} & \multicolumn{1}{c}{LOS}  & \multicolumn{1}{c}{Gender} & \multicolumn{1}{c}{Ethnicity} & \multicolumn{1}{c}{Age}\\
    \cmidrule(lr){1-9} 
    \multicolumn{1}{c}{Unique Entries} & 6,154  & 1,926 & 197 & 985 & 130 & 2 & 41 & 83\\
    \multicolumn{1}{c}{Avg. Entries per Hospitalization ± SD} & 18.248 $\pm$ 21.745 & 6.204 $\pm$ 5.921 & 28.974 $\pm$ 15.906 & 85.182 $\pm$  32.913 & - & - & - & -\\
    \multicolumn{1}{c}{Most Frequent code (Frequency)} & 4019 (19,653) & 3893 (14,991) & B05X (33,501) & N 50971 (35,739) / Un 51219 (34,908)* & 5 (3,404) & M (20,619) & White (25,964) & over 90 (1,236)\\
    \multicolumn{1}{c}{2nd Most Frequent code (Frequency)} & 4280 (14,553) & 9904 (8,423) & N02B (32,359) & N 50868 (35,710) / Un 51221 (34,758) & 6 (3,300) & F (15,307) & Black/African American (3,146) & 62 (835)\\
    \multicolumn{1}{c}{3rd Most Frequent code (Frequency)} & 42731 (12,849) & 966 (8,049) & A02B (32,148) & N 50983 (35,519) / Un 50931 (34,679) & 7 (2,960) & - & Unknown/Not specified (2,294) & 61 (817)\\
    \cmidrule(lr){1-9} 
    
    & \multicolumn{8}{c}{MIMIC-IV Dataset (Events: 224,670, Patients: 119,315)} \\
     \cmidrule(lr){2-9} 
     \multicolumn{1}{c}{} & \multicolumn{1}{c}{Diagnosis} & \multicolumn{1}{c}{Procedures} & \multicolumn{1}{c}{Medications} & \multicolumn{1}{c}{Lab events} & \multicolumn{1}{c}{LOS}  & \multicolumn{1}{c}{Gender} & \multicolumn{1}{c}{Ethnicity} & \multicolumn{1}{c}{Age}\\
    \cmidrule(lr){1-9} 
    \multicolumn{1}{c}{Unique Entries} & 22,601  & 12,563 & 201 & 985 & 160 & 2 & 33 & 85\\
    \multicolumn{1}{c}{Avg. Entries per Hospitalization $\pm$ SD} & 35.673 $\pm$ 61.176 & 6.669 $\pm$ 8.839 & 22.912 $\pm$ 17.138 & 85.182 $\pm$  32.913 & - & - & - & -\\
    \multicolumn{1}{c}{Most Frequent code (Frequency)} & 4019 (135,553) & 3995 (51,030) & B02B (200,084) & N 50971 (187,725) / Un 51221 (175,900)* & 3 (34,294) & F (116,707) & White (147,054) & over 90 (6,530)\\
    \multicolumn{1}{c}{2nd Most Frequent code (Frequency)} & 2724 (111,302) & 3893 (38,333) & A06A (178,330) & N 50983 (184,456) / Un 51279 (175,581) & 2 (31,823) & M (107,963) & Black/African American (25,371) & 64 (5129)\\
    \multicolumn{1}{c}{3rd Most Frequent code (Frequency)} & 4280 (70,627) & 3897 (28,986) & B01A (174,366) & N 50868 (183,669) / Un 51222 (171,181) & 4 (28,141) & - & Unknown/Not specified (9,221) & 66 (5,061)\\
    \toprule
    \end{tabular}
  }
  \label{tab:mimic-iii}
\footnotesize{*In the lab event notation, ``N'' denotes normal, while ``UN'' signifies abnormal.}
\end{table*}

\subsection{Structural correlation among validation sets}
\begin{figure*}[!htbp]
	\centering	
    \includegraphics[width=12.5cm]{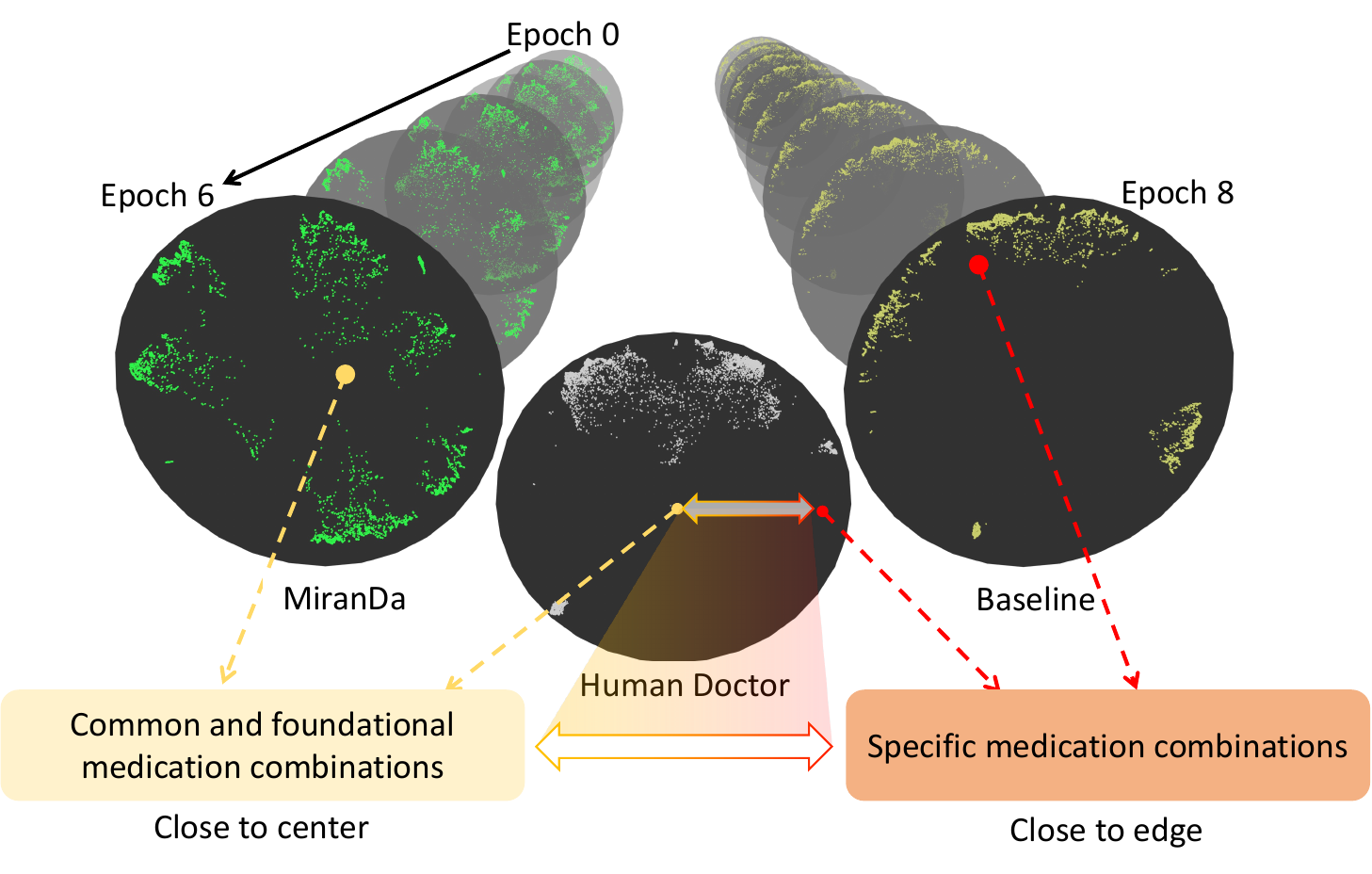}
	  \caption{\textbf{Comparative Analysis of Spatial Positioning in the Poincaré Model for Validation Set during Training}. This figure shows the training trajectory of both \textbf{MiranDa} and the baseline method, and the disc at the center illustrates the actual spatial distribution from human doctors. \textbf{MiranDa} achieved its training in six epochs, compared to the baseline's eight epochs. The centrally positioned disc symbolizes genuine spatial distribution. Each point indicates a medication combination from a patient's hospitalization. The structured data information inherently featured in human doctor decisions is mimicked by \textbf{MiranDa} to replicate this hierarchical structure effectively. In contrast, the baseline method demonstrated a conspicuous absence of a discernable hierarchical structure, reflecting a potential discrepancy in the emulation of human-like decision-making processes. In the utilization of the Poincaré model for organizing or representing data, the position of an object approaches the center of the model, the attributes or characteristics it represents exhibit a higher level of foundational significance and abstraction within the entire structure. This implies that points situated near the center symbolize more fundamental or universal concepts, whereas those further from the center correspond to more concrete or specific instances.}
\label{fig:3d_val}
\end{figure*}

We employed a Poincare model to visualize the spatial positioning of the structural correlation of the validation set, as depicted in Figure \ref{fig:3d_val}. \textbf{MiranDa} mimics the structured medication combinations used in human doctor decisions to replicate this hierarchical structure effectively. In contrast, the baseline method lacks a clear hierarchical structure, suggesting a discrepancy in emulating human-like decision-making.
\subsection{Model performance}

Our initial task involved contrasting the real LOS from records with the ELOS retrieved from patients, aiming to evaluate our retrieval strategy. As depicted in Table \ref{tab:elos}, there is a remarkable similarity between the real LOS (11.087 $\pm$ 0.141) and the ELOS (11.088 $\pm$ 0.141). The difference is minuscule, amounting to approximately 0.001 day (roughly 1.44 minutes). This striking alignment underscores the reliability of our ELOS derivation method for clinical applications and affirms its suitability for causal inferences.
\begin{table}[h]
\centering
\caption{\textbf{Comparison between real LOS and ELOS }}
\setlength{\tabcolsep}{50pt}
\scriptsize

\begin{tabular}{lcc}
\toprule
    & Real LOS$^a$ & ELOS$^b$   \\  \hline
MIMIC-III  &  11.087 $\pm$ 0.141 & 11.088 $\pm$ 0.141 \\
MIMIC-IV  &  6.653 & 6.651 \\
 \bottomrule
\end{tabular}
\\
\raggedright 
\footnotesize{a: The average length of stay of each patient according to the MIMIC database. b: The average length of stay retrieved based on recorded medication combinations from the MIMIC database.}
\label{tab:elos}%
\end{table}
Turning our attention to model performance, \textbf{MiranDa} performs better than baseline, as presented in Table \ref{tab:comparison}. Here, the baseline is the traditional supervised learning model from the Therapeutic Optimization Phase, which has the same architecture as \textbf{MiranDa}. A notable difference was detected in ELOS values: the Baseline yielded 11.137 $\pm$  0.153 for MIMIC III and 6.852 for MIMIC VI, whereas the \textbf{MiranDa} model demonstrated slightly lower figures, at 11.126 $\pm$  0.155 for MIMIC III and 6.844 for MIMIC VI. The ELOS difference in MIMIC III corresponds to enhanced therapeutic effectiveness and potentially improves resource allocation by approximately 15.8 minutes across an aggregate of 150,000 hospitalization events, derived from 30 distinct test datasets. An improvement from the day-level metric suggests that the model can identify and leverage patterns leading to shorter hospital stays, even within the coarse granularity of full days. However, it's crucial to highlight that our model's capability to discern and effect improvements at the day-level metric signifies a potentially even greater impact when considered at the hour-level.

\begin{table*}[htbp]
  \centering
  \caption{\textbf{Comparison of performance between baseline and MiranDa}}
  \resizebox{\linewidth}{!}{
    \large
    \begin{tabular}{ccccccccc}
    \toprule
    
    & \multicolumn{8}{c}{MIMIC-III} \\
    \cmidrule(lr){2-9} 
    
    & \multicolumn{1}{c}{ELOS$\downarrow$} & \multicolumn{1}{c}{ROC-AUC (\%)$\uparrow$} & \multicolumn{1}{c}{PR-AUC (\%)$\uparrow$} & \multicolumn{1}{c}{Jaccard (\%)$\uparrow$} & \multicolumn{1}{c}{Recall (\%)$\uparrow$} & \multicolumn{1}{c}{Precision (\%)$\uparrow$} & \multicolumn{1}{c}{F1 (\%)$\uparrow$} & \multicolumn{1}{c}{DDI (\%)$\downarrow$} \\
    
    \cmidrule(lr){2-9} 
    Baseline & 11.137 $\pm$ 0.153 & 93.953 $\pm$ 0.071 & 76.220 $\pm$ 0.209 & 46.971 $\pm$ 0.553 & 56.419 $\pm$ 1.107 & \textbf{75.703 $\pm$ 0.679} & 62.204 $\pm$ 0.534 & \textbf{5.968 $\pm$ 1.385}\\
    \textbf{MiranDa} & \textbf{11.126 $\pm$ 0.155}$\ast$ & \textbf{93.979 $\pm$ 0.066}$\ast$ & \textbf{76.315 $\pm$ 0.198}$\ast$ & \textbf{47.243 $\pm$ 0.535}$\ast$ & \textbf{56.868 $\pm$ 1.261} & 75.609 $\pm$ 0.980 & \textbf{62.448 $\pm$ 0.515}$\ast$ & 6.027 $\pm$ 0.130\\
    
    \cmidrule(lr){1-9}
    & \multicolumn{8}{c}{MIMIC-IV} \\
    \cmidrule(lr){2-9} 
    
    & \multicolumn{1}{c}{ELOS$\downarrow$} & \multicolumn{1}{c}{ROC-AUC (\%)$\uparrow$} & \multicolumn{1}{c}{PR-AUC (\%)$\uparrow$} & \multicolumn{1}{c}{Jaccard (\%)$\uparrow$} & \multicolumn{1}{c}{Recall (\%)$\uparrow$} & \multicolumn{1}{c}{Precision (\%)$\uparrow$} & \multicolumn{1}{c}{F1 (\%)$\uparrow$} & \multicolumn{1}{c}{DDI (\%)$\downarrow$} \\
    \cmidrule(lr){2-9} 
    Baseline & 6.852 & 93.819 & 71.372 & 40.433 & 47.979 & \textbf{75.083} & 55.380 & 5.988\\
    \textbf{MiranDa} & \textbf{6.844} & \textbf{93.841} & \textbf{71.422} & \textbf{41.089} & \textbf{49.159} & 74.408 & \textbf{56.049} & \textbf{5.923}\\

    \bottomrule
    \end{tabular}
  }
  \label{tab:comparison}
  \raggedright 
  \footnotesize{*Statistical significance is indicated by an asterisk for differences where \(p<0.05\). Abbreviations: ELOS, Estimated Length of Stay in the Intensive Care Unit; ROC-AUC, Receiver Operating Characteristic Area Under Curve; PR-AUC, Precision-Recall Area Under Curve; DDI, drug-drug interaction.}
\end{table*}

\textbf{MiranDa} model also showed a significantly higher score in ROC-AUC (93.979 $\pm$ 0.066), PR-AUC (76.315 $\pm$ 0.198), Jaccard similarity (47.243\% $\pm$ 0.535), recall (56.868\% $\pm$ 1.261), and F1-score (62.448\% $\pm$ 0.515). Our tests on the MIMIC IV data show better performance in terms of ELOS (6.833) and ROC-AUC (93.834\%) metrics as well. These metrics attest to the overall superior performance of the \textbf{MiranDa} regarding prediction accuracy and robustness. Despite a minor increase in DDI from 5.968\% $\pm$ 1.385 for Baseline to 6.027\% $\pm$ 0.130 for our model, the \textbf{MiranDa} exhibited a significant improvement in prediction performance.

\subsection{Descriptive analysis of cluster groups}

\begin{figure*}[!htbp]
	\centering	
    \includegraphics[width=14cm]{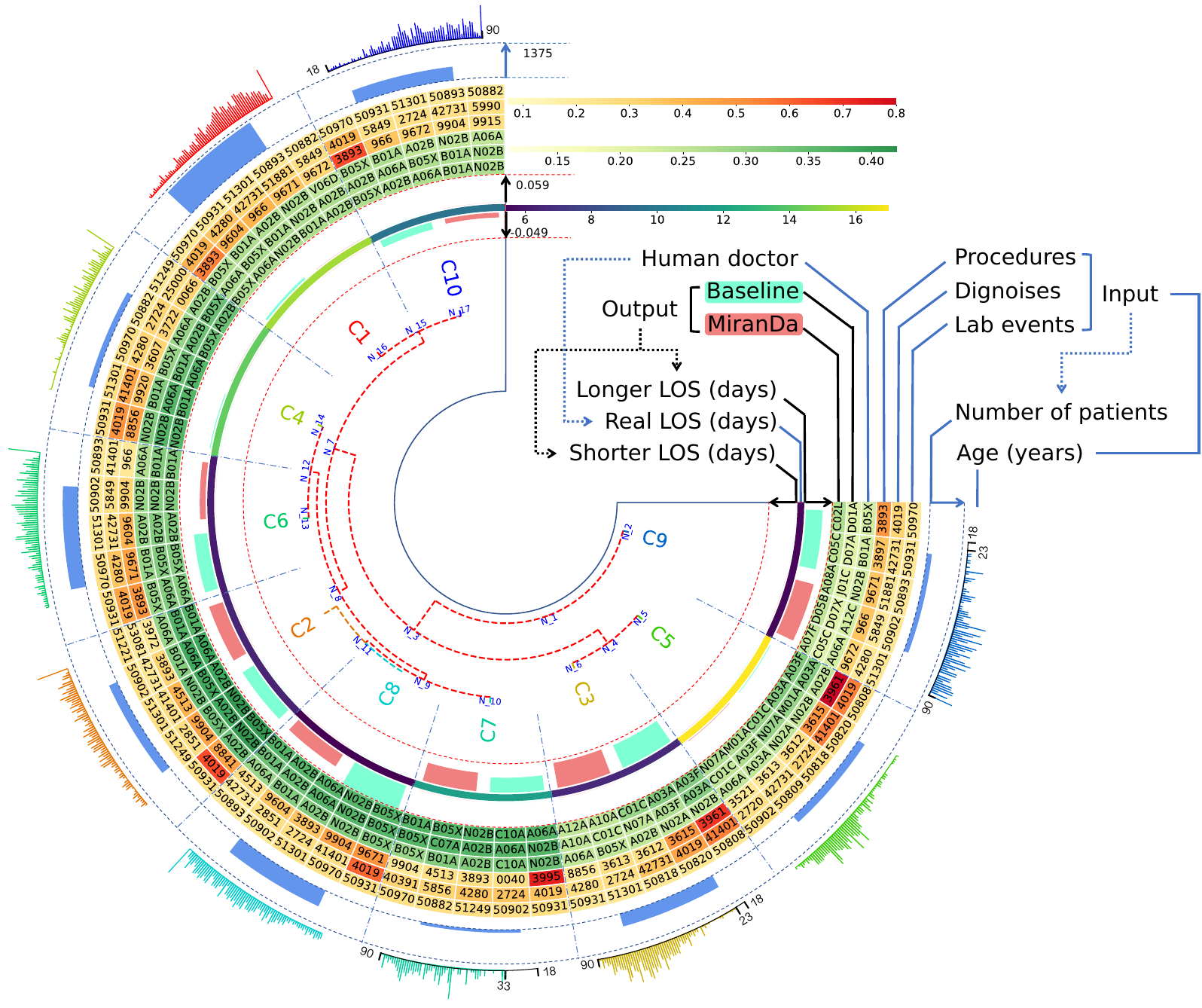}
	  \caption{\textbf{Hierarchical Clustering Dendrogram and Associated Heatmap Analysis}. Starting from the center and moving outward, the first layer illustrates a Hierarchical Clustering Dendrogram with label colors indicating categories within the hyperbolic space and Poincaré model. The next three layers delineate the number of days: fewer than the patient's Length of Stay (LOS), equal to the LOS, and exceeding the LOS, respectively. The subsequent heatmap layer comprises six data sectors: \textbf{MiranDa}, baseline, human doctor, procedure, diagnosis, and lab events. The penultimate layer visualizes the number of individuals within the reconfigured group, and finally, the outermost layer presents the corresponding age distribution. The specifics regarding the actual values can be found in the supplementary data files S1 to S5.}
\label{fig:tfidf}
\end{figure*}

To assess the efficacy of the optimal medication recommendations, we delve deeper into the content within each cluster, as illustrated in Fig. \ref{fig:tfidf}. Our clustering is grounded on \textbf{MiranDa} predictions. Thematic Representations from both baseline and human doctors are constructed using congruent patient hospitalization IDs. From this, themes related to procedures, lab events, diagnostics, and age are subsequently outlined.
\subsection{Procedure-specific attributes of medication combinations}
A detailed examination of the Term Frequency-Inverse Document Frequency (TF-IDF) outcomes highlights the \textbf{MiranDa}'s distinct advantage in term differentiation within clusters. Remarkably, our model often yields higher TF-IDF values compared to the baseline methodologies. For instance, within cluster C6, the term "A06A" (pertaining to drugs for constipation) assigned a TF-IDF value of 0.317 by our model, surpasses the Baseline value of 0.282, and comes close to the Human doctor's 0.312. Such a trend of elevated TF-IDF values under our model is consistent across all clusters. Although there is consistency between the \textbf{MiranDa} and baseline methodologies in terms of the primary terms linked to each cluster, this congruence is absent when comparing the actual TF-IDF values produced by both methods.

The results from human doctors, our reference standard, display a balanced distribution of TF-IDF values across the top terms. Its top term choices consistently appear within the top selections of the \textbf{MiranDa} and Baseline, albeit in different orders and with varying TF-IDF values. Such as the frequent occurrence of certain terms - "A06A" (drugs for constipation), "B05X" (Blood substitutes and perfusion solutions: I.V. solution additives), and "A02B" (Peptic ulcer) - among the top terms across several clusters in both models. This could potentially signal their widespread relevance across various contexts.
A unique trend was observed in category C7, related to other and unspecified hyperlipidemia (2724), congestive heart failure (4280), end-stage renal disease, hypertensive chronic kidney disease (40391), and unspecified essential hypertension (4019). In this category, \textbf{MiranDa} demonstrated a shorter ELOS (-0.0357, n for C7 = 142) than both the baseline (-0.0287) and the Human doctor with the real LOS of 11.8048 days. Interestingly, the Human doctor and our model recommended Lipid modifying agents, plain Pantethine (C10A), for this diagnosis, a prescription noticeably absent from the baseline's top 5 suggestions. Moreover, \textbf{MiranDa} preferred A06A over A02B in its top 5 recommendations.

\subsection{Structural attributes of medication combinations}
\begin{figure*}[htbp]
	\centering	
    \includegraphics[width=16cm]{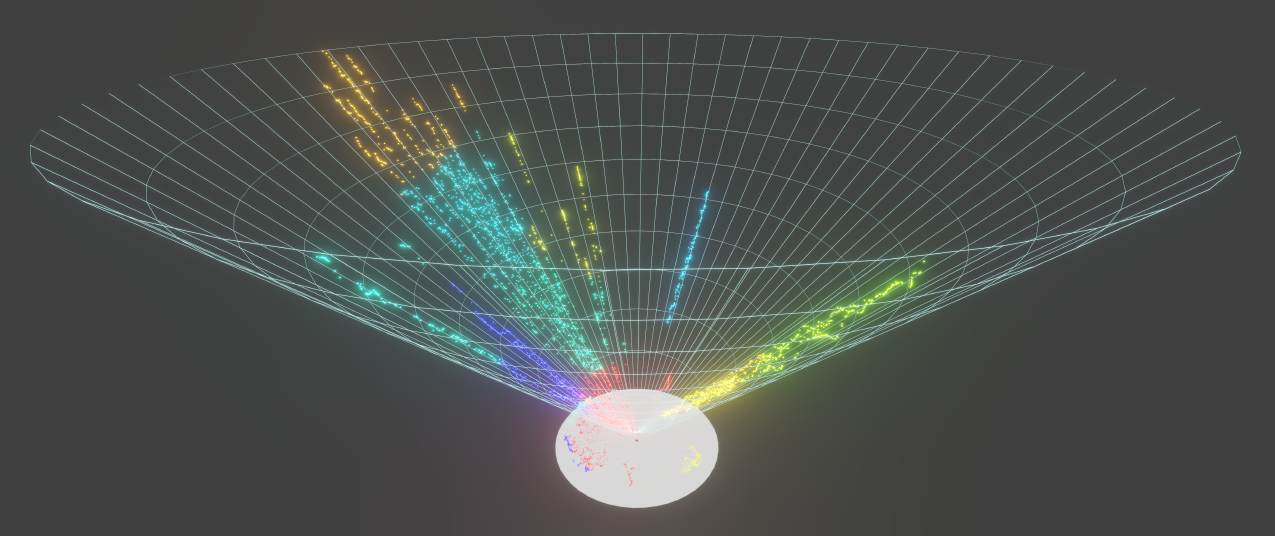}
	  \caption{\textbf{Spatial Distribution of Test Set Data in Hyperbolic Space and The Corresponding Poincaré Model Mapping}. The color gradations across different regions within the figure represent the classification outcomes from k-means analysis. The key detailing color-class correspondence is the same as Fig \ref{fig:tfidf}. For enhanced visual clarity, the Poincaré model depiction has been magnified.}
\label{fig:3d_scatter}
\end{figure*}

As illustrated in Figure \ref{fig:3d_scatter}, the structural attributes of medication combinations can be visualized using \textbf{MiranDa}. This bears similarity to the hierarchical thematic representation presented in Figure \ref{fig:tfidf}. To optimize the representation of spatial relationships, we initially map the data to the upper hemisphere of the hyperbolic space, followed by a transformation into the Poincare model. Upon examination of the results, we discern a robust relational structure. Notably, a category, C1,  with a high degree of universality is observed proximal to the apex of the lower hyperbolic space. Notably, C1 also emerges as the cluster in the final split during clustering. These observations underscore a potential focal point of commonality within the data, underscoring the efficacy of our \textbf{MiranDa} algorithm in identifying structured patterns within complex datasets.

\begin{figure*}[!htbp]
	\centering	
    \includegraphics[width=16cm]{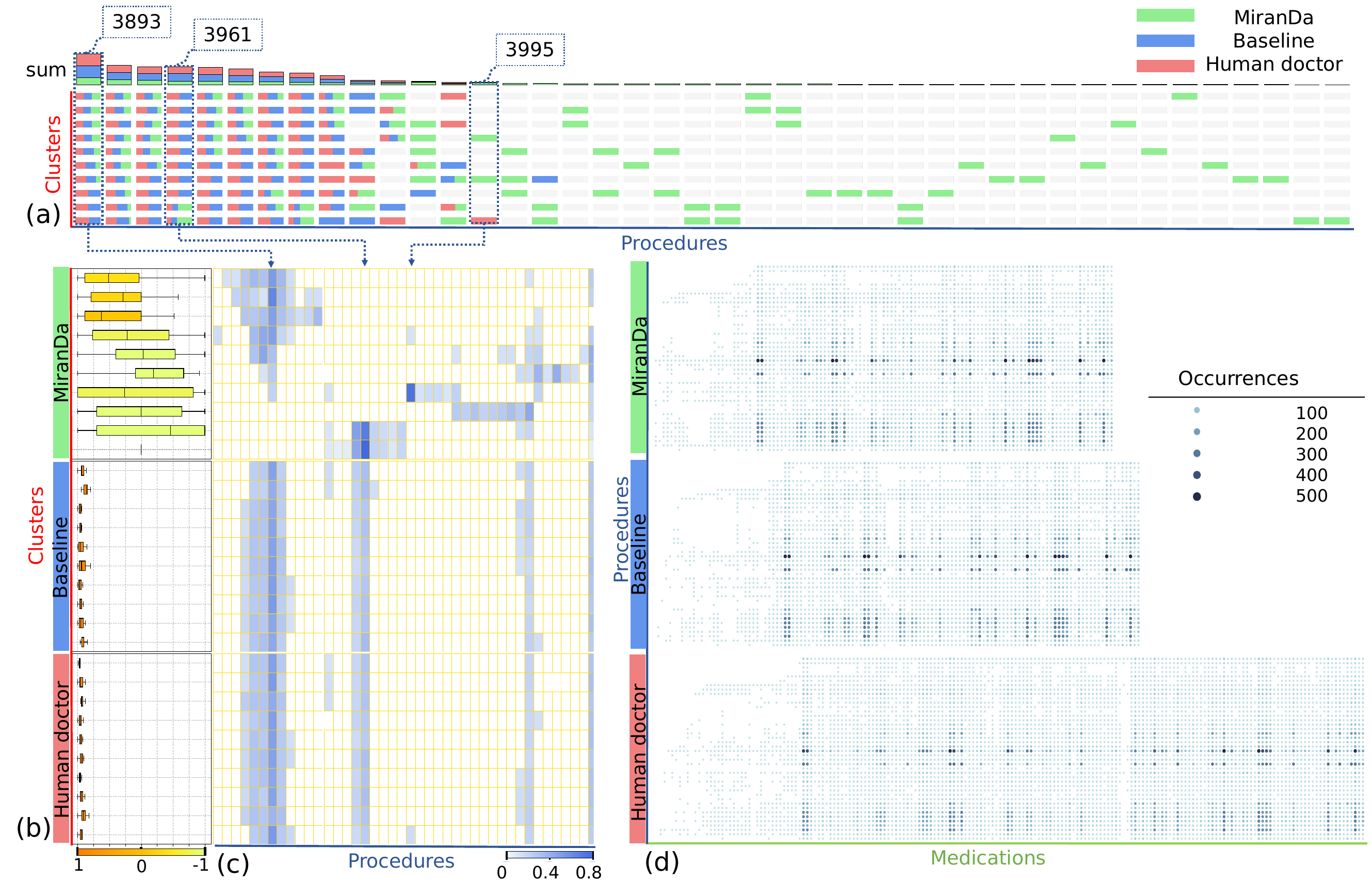}
	  \caption{\textbf{Comparative Analysis in Medication-Procedure Clustering}. (a) Reassignment of Clusters Through \textbf{MiranDa} Clustering Model: This graph plots procedures horizontally against ten clusters (with cross-data clustering) for our model, baseline, and human doctors. In each grid, colored sections represent similar procedures per class, gauged by the TF-IDF metric. (b) Cluster Correlation Box Plot: This shows a matrix of correlations between procedures and clusters (with inherent data clustering) from our model, baseline, and human doctors. Clusters of our model exhibit a wider correlation range, unlike the baseline and human doctors that trend near unity. (c) Heatmap Analysis: This highlights variations in procedure information within clusters of our model, indicating a broader procedure range than the baseline. (d) Medication-Procedure Co-occurrence Plot: This diagram depicts the frequency of medication-procedure pairs. The ground truth shows more consistent pairings, while our model has fewer concurrent medications, hinting at its reliability. This study infers fewer concurrent medications might enhance medical outcomes, given \textbf{MiranDa}'s edge over the baseline.}
\label{fig:proc_tfidf}
\end{figure*}

With structured medication information, potential procedure targeting emerges with different clusters. In detail, the medication combinations of \textbf{MiranDa} are more "procedure-specific" due to the rationalization enhanced by the possible combinations of drugs in the RL phase, while at the same time, \textbf{MiranDa} uses fewer medications to achieve a higher model performance. Here, we clustered the medications and then looked at the distribution of procedures according to category, as shown in Figure \ref{fig:proc_tfidf}. 
\paragraph{Assignment of clusters through \textbf{MiranDa} clustering model}
To extend our analytical scope, we adopt the \textbf{MiranDa}'s clustering model to assign clusters for both the baseline and Human doctor, as Figure \ref{fig:proc_tfidf}(a). We established a clustering model using the medication combinations from \textbf{MiranDa} and then applied this model to cluster the baseline and human doctor. This methodology is pursued with the understanding that our model's clustering may reveal unique or nuanced structures within the dataset that might not be evident when employing traditional methods or Human doctor algorithms. The results showed that even with the medication combination structure of our model, human doctors and baselines still did not appear to be targeted to procedures. On the other hand, our model embraces an array of unique procedures. Such a finding, which harmonizes with previously reported results, underscores the capacity of \textbf{MiranDa} to refine medication combinations by leveraging exhaustive procedural information.

\paragraph{Procedural data elucidation}

The inherent data clustering approach, each dataset was clustered independently based on its own medication combinations, reveals an intriguing divergence when cross-group correlations are considered, as shown in Figure \ref{fig:proc_tfidf}(b). Unlike the baseline and the Human doctor, where average cluster correlation values consistently align close to one, a wider distribution of correlation values is seen with clusters generated by \textbf{MiranDa}. Further, as shown in Figure \ref{fig:proc_tfidf}(c), our model displays a persistent predilection for diversifying procedures across clusters, thereby underlining its robust capability to focus on procedural information. This finding illustrates our model has a more dynamic variability in procedure-cluster relationships—an attribute unseen with baseline and human doctors. 

\paragraph{Medication-procedures co-occurrence relationship}

Delving into a co-occurrence plot of medications and procedures, with medications mapped on the x-axis and procedures on the y-axis, the real labels encompass a greater number of medications for the same procedures compared to \textbf{MiranDa}, which maintains a minimal medication list, as shown in Figure \ref{fig:proc_tfidf}(d). This underlines the ability to yield improved medical outcomes via a leaner pharmacological approach potentially, provided \textbf{MiranDa} outperforms the baseline.

\section{Discussion}

Our research introduces a causal inference-based paradigm, which combines RL and supervised learning, optimally suited for a wide range of deep learning tasks. This approach allows to generate counterfactual outcomes, informed by original training data, which dynamically evaluates the validity and efficacy of predictions within real-world contexts. This paradigm aims to achieve superior predictive outcomes and mitigate inherent dataset flaws. Further, we illustrated this concept by developing \textbf{MiranDa}, a model that uses transformers as its core structure and simulates the learning and progression of physicians from school to medication via RL with the start point from supervised learning. In the context of our model, the agent represents the clinical practitioner, and the environment is the patient's health state, with available actions being the range of possible pharmaceutical interventions. The reward function, defined by the mean hospitalization duration, guides the agent towards actions that minimize this duration, effectively leading to optimal medications. \textbf{MiranDa} provides robust and efficient medication combination recommendations, showing significantly improved performance across ELOS, ROC-AUC, PR-AUC, Jaccard and F1 metrics. Moreover, our model promotes structural and "procedure-specific" attributes of medication combinations.

In assessing the comparative performance of the Baseline and \textbf{MiranDa} as detailed in Table \ref{tab:comparison}, our findings highlight a marked improvement achieved by \textbf{MiranDa} in terms of ELOS and prediction accuracy. This superiority signifies our model's precise identification of correct medication combinations, offering enhanced efficacy. However, \textbf{MiranDa} has a higher DDI rate than the baseline in the MIMIC III database, which corroborates the viewpoint we posited in the background, that an excessive reliance on DDI for optimization may not significantly benefit the ultimate medical outcomes.

Importantly, the medication combinations made by \textbf{MiranDa} closely parallel those of human doctors. For example, C7 highlights complex medical conditions, including hyperlipidemia, congestive heart failure, and end-stage renal disease, these conditions amplify the requirement for constipation medications. However, the baseline model fails to recognize these intricate relationships. First and foremost, managing opioid analgesics for pain relief in terminal diseases is a pivotal consideration. Opioids (N02), which exhibit the highest prevalence in Category C7 for human doctors and \textbf{MiranDa}, are well-known for inducing constipation due to their suppressive impact on gastrointestinal motility \cite{zollner2007opioids}. Besides, numerous medications used in heart and blood pressure, can trigger constipation through diverse pharmacological mechanisms such as altering intestinal motility or gut microbiota \cite{weersma2020interaction}. 
Secondly, hemodialysis, a prevalent treatment for renal disorders, can exacerbate constipation. This stems from the recommended fluid intake restrictions for these patients and potential changes in gut microbiota attributable to renal conditions and the dialysis procedure. Moreover, the actual LOS for patients in C7 averages 11.8048 days. Such an extended duration may imply reduced physical activity. Coupled with specific low-fiber dietary prescriptions, this can heighten constipation risks, necessitating the administration of laxatives. Regrettably, the baseline model seems oblivious to these multifaceted interactions, resulting in subpar medication suggestions. This underscores our model's performance to comprehend complex scenarios, approximating the understanding of human doctors.

Meanwhile, a salient structural attribute of medication combinations is captured from \textbf{MiranDa}, which indicates the progressive relationship between medical prescriptions. Within the confines of the Poincaré model, validation sets were presented that were similar to those encountered by human doctors during the learning process. Furthermore, we visualized the \textbf{MiranDa} model's outputs from the test set, which exhibited comprehensive and fundamental structural categorizations in hyperbolic space, with a more significant number of predicted samples assigned near the central region, which reflects the progressive relationship between medical prescriptions. To exemplify, the treatment of hypertension necessitates a broad array of antihypertensive medications, like diuretics, beta-blockers, or calcium channel blockers, as the initial line of management. Over time, if these therapies don't lead to adequate blood pressure control, more specialized treatments such as aldosterone antagonists or alpha-blockers can be applied \cite{al2022hypertension}. Understanding the function and derivations of medication combinations is crucial for optimizing therapeutic plans, and this is where our method shines.

Specifically, the medication combinations employed by \textbf{MiranDa} exhibit a higher degree of specificity towards particular procedures. We first conducted clustering on the medication recommendations from human doctors, the baseline and our model, respectively. Our analysis revealed that only the procedures of \textbf{MiranDa} exhibited pronounced procedure-specific. Subsequently, leveraging the clustering structure derived from our model, we re-clustered the medication outcomes for human doctors, the baseline and ours. Consistently, our model demonstrated procedure-specific attributes of medication combinations. Furthermore, when considering individual medications, \textbf{MiranDa} deploys fewer medications to achieve superior model performance.

The marked efficiency of \textbf{MiranDa} can be largely attributed to its success in causal inference, amalgamating supervised learning with RL. Supervised learning (from the Evidence-based Training Phase) ensures a good foundational understanding of the data, and RL (from the Therapeutic Optimization Phase) allows the model to refine its recommendations based on the counterfactual outcomes. Traditional models, especially in healthcare, often follow a supervised learning approach where the model learns from labeled data and makes predictions based on this historical data \cite{an2023comprehensive}. On the one hand, supervised learning could ensure that initial treatment recommendations are based on proven medical practices and speed up the learning process in the Therapeutic Optimization Phase, which could avoid instability from directly applying RL with a randomly initialized state. In medication, it is more like incorporating expert knowledge through supervised learning, which helps ensure that the model's initial behavior is medically sound. In a medical setting, randomly experimenting with treatments (as a purely RL approach might do) can be unsafe. On the other hand, supervised learning also avoids favoring treatments for milder conditions to reduce ICU days by rooting its learning in a wide array of patient data, ensuring that it doesn’t provide overly conservative or simplified recommendations. Therefore, we ensure the initial policy has some sensible behavior that the RL agent can refine further.

Further, the bidirectional constraint ($retrieval^p$ and $retrieval^d$) system could be regarded as constraint satisfaction problems (CSPs). CSPs are essentially problems defined as a set of objects whose state must satisfy several constraints or restrictions \cite{miltzow2022classifying}, such as in a Sudoku or a chess problem. However, while the \textbf{MiranDa} method shares these commonalities with CSPs, $retrieval^p$ and $retrieval^d$ are not traditional CSPs, but rather complex problems in RL and decision-making. The $retrieval^p$ and $retrieval^d$ functions, designed to handle the constraints set by the patient's unique conditions and situations, mimic this principle and apply it to medication recommendations. The constraints act as an extra dimension of information that could enhance the accuracy of the model's recommendations. From the medical standpoint, applying CSPs to patient recommendation procedures appears to be an innovative approach. The constraints can incorporate various factors unique to each patient's case, such as their specific medical condition, age, lifestyle, genetic predisposition, etc. This results in the model's exceptional capability to capture, process, and utilize structural attributes of medicine combinations for recommending accurate and effective medication combinations.

In detail, the \textbf{MiranDa} model creates a more nuanced and informed approach $retrieval^p$ to medication recommendation through the incorporation of patient-specific factors and similarity retrieval, which ensures robust exploration of potential medication combinations under multiple constraints, inspired by patient similarity analysis \cite{patientsimilarityanalysis}, wherein identifying patient trajectories akin to a reference patient aids in predicting the clinical outcomes for said patient. Firstly, the action space represents the set of all possible actions an agent can take, the richness of the action space is a critical factor in the complexity and flexibility of the model \cite{zhu2021overview}. Essentially, $retrieval^p$ creates a set of auxiliary action spaces that may encompass more specific or nuanced actions that the model could take, leading to a more diversified approach to decision-making. The $retrieval^p$ can be likened to providing the model more tools or options to work with while trying to solve the optimization problem, to lead the model to begin with a richer understanding of the action space. In real-world medical scenarios, the ability to consider various possible actions is crucial \cite{walker2022program}. Different patients exhibit varying responses to various treatments, and a more extensive action space allows the model to take these differences into account and be more adaptable to various scenarios. For instance, if certain medications are unavailable or contra-indicated for a patient, having a broader action space allows the model to provide alternative recommendations. 
Secondly, procedures sharing similar medications should significantly improve the alignment between procedures and medications through iterative RL exploration phases. This process is expected to guide the model towards excluding medications potentially extending LOS, while consistently reinforcing an optimized medication combination strategy.
Lastly, a potential benefit is that the RL process from similar patients relies on the aggregated historical data from similar patient treatments. Therefore, even infrequent circumstances or unique combinations of symptoms could be effectively addressed. 
Consequently, we believe that the $retrieval^p$ is the reason that, the \textbf{MiranDa} model showcases "procedure-specific" attributes, which exhibit a more targeted correlation with medications. 

Indeed, the function $retrieval^d$ acts as a filtering mechanism for the auxiliary action space $\mathcal{B}_{pi}$. By selecting the top $\mathcal{K}$ vectors in $\mathcal{B}_{pi}$ that exhibit the highest cosine similarity with ${\alpha}_i$ and surpass a certain threshold ${\phi}$, the model ensures that the selected actions ${\alpha}_i$ align well with the current state. It also allows for greater consistency in the generated actions, as the chosen vectors will have a high degree of similarity with ${\alpha}_i$, thus ensuring a logical progression in the sequence of actions. This approach was spurred by concepts such as medication repositioning \cite{drugrepositioning}. In a medical context, by constraining the action space in a way that retains those actions most closely aligned with the patient's current state, the model can focus on the most relevant treatments \cite{nanayakkara2022unifying,raghu2017continuous}. This kind of focused approach is much akin to a doctor refining a differential diagnosis based on the medications and test results, thereby eliminating less relevant possibilities and zeroing in on the most probable causes \cite{scali2015primary}. 

\textbf{MiranDa} utilized an RL optimization method characterized by a two-tier approach, combining an $\lambda$ gradient update with sequential $\epsilon (1-\lambda)$ updates, each augmented by logarithmic perturbations. Of course, we think the most crucial in the optimization section, is the condition $\hat{\mathcal{H}}_B$, which asks the model to enter and quit the Therapeutic Optimization Phase. 
Optimization aims to minimize the loss by gradient \cite{lecun2015deep}. In our scenario, this perturbed objective function consists of both the Binary Cross Entropy (BCE) and $\hat{\mathcal{H}}_B$ components. Typically, the BCE loss directs model parameter updates toward decreasing prediction errors, effectively guiding the model toward areas of higher predictive accuracy. Thus, without other factors, the BCE loss would primarily orient the gradient in this direction. But, the component ELOS is integrated into the RL perturbation, $\mathcal{P}_{RL}(\theta)$, to encode our objective of minimizing mean hospitalization duration. However, since $\hat{\mathcal{H}}_B$ is not a differentiable function, it cannot directly impact the gradient, instead, $\hat{\mathcal{H}}_B$ influences the gradient by introducing a form of perturbation to the original BCE loss. This perturbation affects the gradient's magnitude and direction, hence nudging the model towards parameters that minimize $\hat{\mathcal{H}}_B$ \cite{minervini2023adaptive}. Firstly, the perturbation influences Gradient Direction: Given the presence of the $\hat{\mathcal{H}}_B$ perturbation, the gradient's direction may alter \cite{yu2019gradient}. If $\hat{\mathcal{H}}_B$ exhibits significant variability, the gradient direction may shift substantially correspondingly. This implies that the model might need to strike a balance between locating the direction that minimizes prediction errors and that which minimizes $\hat{\mathcal{H}}_B$. While this could enhance the model's explorative capabilities, it could also potentially destabilize the learning process. Secondly, perturbation could impact gradient magnitude: The size of $\hat{\mathcal{H}}_B$ also affects the gradient's magnitude \cite{agarwal2021towards,yu2019gradient}. A large $\hat{\mathcal{H}}_B$ value would introduce a correspondingly large perturbation, thereby increasing the gradient magnitude \cite{yu2019gradient}. This could intensify parameter updates, further enhancing the model's exploratory tendencies but could also exacerbate learning process instability. Lastly, influence on Parameter Updates: The introduction of calculated logarithmic perturbations serves as a tool to circumvent regions of low gradient, potentially aiding in the escape from saddle points and local minima \cite{hazan2016perturbations}. This deliberate disturbance of the optimization path may enhance convergence to more globally optimal solutions. Under the combined effects of BCE and $\hat{\mathcal{H}}_B$, the direction of model parameter updates may deviate from merely minimizing prediction errors. This could lead the model to explore more extensive regions in parameter space, seeking solutions that cater to both predictive accuracy and long-term goals. However, this could also introduce model complexity, necessitating more delicate hyperparameter tuning to locate the optimal solution. In summary, introducing the $\hat{\mathcal{H}}_B$ perturbation seeks to balance the model's predictive accuracy and long-term objectives for better performance.


Our research also reveals a potential direction. The \textbf{MiranDa} algorithm could be designed to address a variety of tasks. By adjusting the reward function and the state variables, it could be tailored to different problems, potentially expanding the scope of problems that language models can solve. By combining two types of learning, this approach might yield performance improvements over using just one or the other. This is especially true in cases where a model needs to balance different objectives, such as accuracy, creativity, or novelty. The ability to dynamically adapt the learning strategy could potentially help the model find a better balance between these conflicting goals, leading to better overall performance. With its unique blend of supervised and reinforcement learning and the dynamic switch between these modes, future language models could be much more responsive to the data they encounter. They could make better use of feedback, adjusting their training regime based on the results they are achieving. This could mean more efficient learning, quicker adaptation to new data or scenarios, and ultimately better performance. 


While we've made strides in our research, challenges persist. Firstly, causal inferences from observational studies are complicated by common causes (confounding bias) and selection on common effects (selection bias), which remain in our study. Secondly, our results from both the baseline and \textbf{MiranDa} indicate a notable distinction between AI and human doctors: the complexity of medication. During training, models lean towards frequent labels and fewer to reduce loss. This condition may be a primary barrier to deep learning's efficacy in medicine. Thirdly, our model has not achieved the performance level of human doctors. We continue to maintain that, as of now, the role of artificial intelligence models remains that of an auxiliary to human doctors. We do not believe that AI can achieve a level of treatment surpassing that of doctors merely through limited records. Future endeavors may require the employment of multimodal large language models and the utilization of all available medical records for practical application. Besides, a potential drawback of the method lies in its computational overhead. While we introduce ELOS to guide model updates, it cannot directly update parameters using loss derivation which indicates lower efficiency. The need to recompute loss and gradients, coupled with calculating the log-based perturbations, might make this approach computationally intensive. 

\section{Conclusions}

In conclusion, we allow the "counterfactual outcomes" to be visible, the benefits not only guide parameter updates directly but also directly modify our decisions based on the real-world impact of our decisions. Our model enhanced the efficacy of AI-driven medication suggestions while addressing the limitations inherent in both datasets and current AI models. Our approach provides a significant contribution to introducing a versatile learning paradigm that can be adapted for various deep learning tasks, extending its applicability beyond the medical domain. Additionally, this research has pioneered a novel simulation that emulates the learning trajectory of physicians from their academic phase to medical practice, exhibiting a marked improvement over existing benchmarks. A pivotal advancement in this research is introducing ELOS as counterfactual outcomes. This system is based on real patient data, diverging from the typical algorithm-driven predictions. Lastly, the \textbf{MiranDa} showcases its capability to discern Structural attributes of medication combinations, offering insights into procedure-specific attributes. This underscores a refined understanding of the intricate interplay between medications and medical procedures.
\\
\newpage
\section*{List of Supplementary Materials}
Terms 1-4\\
Data files S1 to S5

\section*{Author contributions}
Conceptualization: ZW\\
Methodology: ZW, XL\\
Investigation: ZW\\
Visualization: ZW, XL\\
Project administration: ZW, RN\\
Supervision: RN\\
Writing – original draft: ZW\\
Writing – review and editing: HM, RN, XL 

\section*{Data and materials availability}

All input data have been accessed through credential verification from public sources, including MIMIC III database (https://physionet.org/content/mimiciii/1.4/) and MIMIC IV database (https://physionet.org/content/mimiciv/0.4/). Data of descriptive analysis of cluster groups are available in the supplementary materials.
\bibliographystyle{unsrt}  
\bibliography{references}

\begin{thebibliography}{10}

\bibitem{gooddrug}
Nick Barber.
\newblock What constitutes good prescribing?
\newblock {\em Bmj}, 310(6984):923--925, 1995.

\bibitem{mcleod2013methodological}
Monsey~Chan McLeod, Nick Barber, and Bryony~Dean Franklin.
\newblock Methodological variations and their effects on reported medication administration error rates.
\newblock {\em BMJ quality \& safety}, 22(4):278--289, 2013.

\bibitem{ADE1}
Lucian~L Leape, David~W Bates, David~J Cullen, Jeffrey Cooper, Harold~J Demonaco, Theresa Gallivan, Robert Hallisey, Jeanette Ives, Nan Laird, Glenn Laffel, et~al.
\newblock Systems analysis of adverse drug events.
\newblock {\em Jama}, 274(1):35--43, 1995.

\bibitem{ADE2}
Lucian~L Leape and Donald~M Berwick.
\newblock Five years after to err is human: what have we learned?
\newblock {\em Jama}, 293(19):2384--2390, 2005.

\bibitem{drugrelatedproblem}
Noe Garin, Nuria Sole, Beatriz Lucas, Laia Matas, Desiree Moras, Ana Rodrigo-Troyano, Laura Gras-Martin, and Nuria Fonts.
\newblock Drug related problems in clinical practice: a cross-sectional study on their prevalence, risk factors and associated pharmaceutical interventions.
\newblock {\em Scientific reports}, 11(1):1--11, 2021.

\bibitem{ADEreason2}
Sarah~P Slight, Clare~L Tolley, David~W Bates, Rachel Fraser, Theophile Bigirumurame, Adetayo Kasim, Konstantinos Balaskonis, Steven Narrie, Andrew Heed, E~John Orav, et~al.
\newblock Medication errors and adverse drug events in a uk hospital during the optimisation of electronic prescriptions: a prospective observational study.
\newblock {\em The Lancet Digital Health}, 1(8):e403--e412, 2019.

\bibitem{ade3}
Martin~A Makary and Michael Daniel.
\newblock Medical error—the third leading cause of death in the us.
\newblock {\em Bmj}, 353, 2016.

\bibitem{ade4}
D~Formica, J~Sultana, PM~Cutroneo, S~Lucchesi, R~Angelica, S~Crisafulli, Y~Ingrasciotta, Francesco Salvo, E~Spina, and G~Trifir{\`o}.
\newblock The economic burden of preventable adverse drug reactions: a systematic review of observational studies.
\newblock {\em Expert opinion on drug safety}, 17(7):681--695, 2018.

\bibitem{ade5}
Dianna Wolfe, Fatemeh Yazdi, Salmaan Kanji, Lisa Burry, Andrew Beck, Claire Butler, Leila Esmaeilisaraji, Candyce Hamel, Mona Hersi, Becky Skidmore, et~al.
\newblock Incidence, causes, and consequences of preventable adverse drug reactions occurring in inpatients: a systematic review of systematic reviews.
\newblock {\em PloS one}, 13(10):e0205426, 2018.

\bibitem{aded1}
Greene Shepherd, Philip Mohorn, Kristina Yacoub, and Dianne~Williams May.
\newblock Adverse drug reaction deaths reported in united states vital statistics, 1999-2006.
\newblock {\em Annals of Pharmacotherapy}, 46(2):169--175, 2012.

\bibitem{aded2}
Leonard~J Paulozzi and Yongli Xi.
\newblock Recent changes in drug poisoning mortality in the united states by urban--rural status and by drug type.
\newblock {\em Pharmacoepidemiology and drug safety}, 17(10):997--1005, 2008.

\bibitem{aded3}
Wu-Chien Chien, Jin-Ding Lin, Ching-Huang Lai, Chi-Hsiang Chung, and Yu-Chen Hung.
\newblock Trends in poisoning hospitalization and mortality in taiwan, 1999-2008: a retrospective analysis.
\newblock {\em BMC Public Health}, 11:1--8, 2011.

\bibitem{aded4}
Peter~A Chyka.
\newblock How many deaths occur annually from adverse drug reactions in the united states?
\newblock {\em The American journal of medicine}, 109(2):122--130, 2000.

\bibitem{goodwin2018association}
James~S Goodwin, Habeeb Salameh, Jie Zhou, Siddhartha Singh, Yong-Fang Kuo, and Ann~B Nattinger.
\newblock Association of hospitalist years of experience with mortality in the hospitalized medicare population.
\newblock {\em JAMA internal medicine}, 178(2):196--203, 2018.

\bibitem{bastian2010seventy}
Hilda Bastian, Paul Glasziou, and Iain Chalmers.
\newblock Seventy-five trials and eleven systematic reviews a day: how will we ever keep up?
\newblock {\em PLoS medicine}, 7(9):e1000326, 2010.

\bibitem{electronicprescribing}
Rainu Kaushal, Lisa~M Kern, Yolanda Barr{\'o}n, Jill Quaresimo, and Erika~L Abramson.
\newblock Electronic prescribing improves medication safety in community-based office practices.
\newblock {\em Journal of general internal medicine}, 25:530--536, 2010.

\bibitem{doctorai}
Edward Choi, Mohammad~Taha Bahadori, Andy Schuetz, Walter~F Stewart, and Jimeng Sun.
\newblock Doctor ai: Predicting clinical events via recurrent neural networks.
\newblock In {\em Machine learning for healthcare conference}, pages 301--318. PMLR, 2016.

\bibitem{micron}
Chaoqi Yang, Cao Xiao, Lucas Glass, and Jimeng Sun.
\newblock Change matters: Medication change prediction with recurrent residual networks.
\newblock {\em arXiv preprint arXiv:2105.01876}, 2021.

\bibitem{medicareai}
Deyin Liu, Yuanbo~Lin Wu, Xue Li, and Lin Qi.
\newblock Medi-care ai: Predicting medications from billing codes via robust recurrent neural networks.
\newblock {\em Neural Networks}, 124:109--116, 2020.

\bibitem{retain}
Edward Choi, Mohammad~Taha Bahadori, Jimeng Sun, Joshua Kulas, Andy Schuetz, and Walter Stewart.
\newblock Retain: An interpretable predictive model for healthcare using reverse time attention mechanism.
\newblock {\em Advances in neural information processing systems}, 29, 2016.

\bibitem{leap}
Yutao Zhang, Robert Chen, Jie Tang, Walter~F Stewart, and Jimeng Sun.
\newblock Leap: learning to prescribe effective and safe treatment combinations for multimorbidity.
\newblock In {\em proceedings of the 23rd ACM SIGKDD international conference on knowledge Discovery and data Mining}, pages 1315--1324, 2017.

\bibitem{gamenet}
Junyuan Shang, Cao Xiao, Tengfei Ma, Hongyan Li, and Jimeng Sun.
\newblock Gamenet: Graph augmented memory networks for recommending medication combination.
\newblock In {\em proceedings of the AAAI Conference on Artificial Intelligence}, volume~33, pages 1126--1133, 2019.

\bibitem{safedrug}
Chaoqi Yang, Cao Xiao, Fenglong Ma, Lucas Glass, and Jimeng Sun.
\newblock Safedrug: Dual molecular graph encoders for recommending effective and safe drug combinations.
\newblock {\em arXiv preprint arXiv:2105.02711}, 2021.

\bibitem{csedrug}
Jialun Wu, Buyue Qian, Yang Li, Zeyu Gao, Meizhi Ju, Yifan Yang, Yefeng Zheng, Tieliang Gong, Chen Li, and Xianli Zhang.
\newblock Leveraging multiple types of domain knowledge for safe and effective drug recommendation.
\newblock In {\em Proceedings of the 31st ACM international conference on information \& knowledge management}, pages 2169--2178, 2022.

\bibitem{pccnet}
Ruobing Li, Jian Wang, Hongfei Lin, Yuan Lin, Huiyi Lu, and Zhihao Yang.
\newblock Patient condition change network for safe medication recommendation.
\newblock In {\em 2022 IEEE International Conference on Bioinformatics and Biomedicine (BIBM)}, pages 1046--1051. IEEE, 2022.

\bibitem{SARMR}
Yanda Wang, Weitong Chen, Dechang Pi, Lin Yue, Sen Wang, and Miao Xu.
\newblock Self-supervised adversarial distribution regularization for medication recommendation.
\newblock In {\em IJCAI}, pages 3134--3140, 2021.

\bibitem{grasp}
Chaohe Zhang, Xin Gao, Liantao Ma, Yasha Wang, Jiangtao Wang, and Wen Tang.
\newblock Grasp: generic framework for health status representation learning based on incorporating knowledge from similar patients.
\newblock In {\em Proceedings of the AAAI conference on artificial intelligence}, volume~35, pages 715--723, 2021.

\bibitem{rldrug1}
Aniruddh Raghu, Matthieu Komorowski, Imran Ahmed, Leo Celi, Peter Szolovits, and Marzyeh Ghassemi.
\newblock Deep reinforcement learning for sepsis treatment.
\newblock {\em arXiv preprint arXiv:1711.09602}, 2017.

\bibitem{rldrug2}
Wei-Hung Weng, Mingwu Gao, Ze~He, Susu Yan, and Peter Szolovits.
\newblock Representation and reinforcement learning for personalized glycemic control in septic patients.
\newblock {\em arXiv preprint arXiv:1712.00654}, 2017.

\bibitem{rldrug3}
Lu~Wang, Wei Zhang, Xiaofeng He, and Hongyuan Zha.
\newblock Supervised reinforcement learning with recurrent neural network for dynamic treatment recommendation.
\newblock In {\em Proceedings of the 24th ACM SIGKDD international conference on knowledge discovery \& data mining}, pages 2447--2456, 2018.

\bibitem{dataseterrors}
Curtis~G Northcutt, Anish Athalye, and Jonas Mueller.
\newblock Pervasive label errors in test sets destabilize machine learning benchmarks.
\newblock {\em arXiv preprint arXiv:2103.14749}, 2021.

\bibitem{drugerror1}
Christopher~M Wittich, Christopher~M Burkle, and William~L Lanier.
\newblock Medication errors: an overview for clinicians.
\newblock In {\em Mayo Clinic Proceedings}, volume~89, pages 1116--1125. Elsevier, 2014.

\bibitem{drugerror2}
MR~Chen and HF~Wang.
\newblock The reason and prevention of hospital medication errors.
\newblock {\em Practical Journal of Clinical Medicine}, 4, 2013.

\bibitem{drugreportincident}
David~J Cullen, David~W Bates, Stephen~D Small, Jeffrey~B Cooper, A~Roberta Nemeskal, and Lucian~L Leape.
\newblock The incident reporting system does not detect adverse drug events: a problem for quality improvement.
\newblock {\em The Joint Commission journal on quality improvement}, 21(10):541--548, 1995.

\bibitem{unnessaryoperation}
J~Bruce Moseley, Kimberly O'malley, Nancy~J Petersen, Terri~J Menke, Baruch~A Brody, David~H Kuykendall, John~C Hollingsworth, Carol~M Ashton, and Nelda~P Wray.
\newblock A controlled trial of arthroscopic surgery for osteoarthritis of the knee.
\newblock {\em New England Journal of Medicine}, 347(2):81--88, 2002.

\bibitem{madian2012relating}
Ashraf~G Madian, Heather~E Wheeler, Richard~Baker Jones, and M~Eileen Dolan.
\newblock Relating human genetic variation to variation in drug responses.
\newblock {\em Trends in genetics}, 28(10):487--495, 2012.

\bibitem{evans2001pharmacogenomics}
William~E Evans and Julie~A Johnson.
\newblock Pharmacogenomics: the inherited basis for interindividual differences in drug response.
\newblock {\em Annual review of genomics and human genetics}, 2(1):9--39, 2001.

\bibitem{hernan2019second}
Miguel~A Hern{\'a}n, John Hsu, and Brian Healy.
\newblock A second chance to get causal inference right: a classification of data science tasks.
\newblock {\em Chance}, 32(1):42--49, 2019.

\bibitem{pearl2009causal}
Judea Pearl.
\newblock {Causal inference in statistics: An overview}.
\newblock {\em Statistics Surveys}, 3(none):96 -- 146, 2009.

\bibitem{johansson2016learning}
Fredrik Johansson, Uri Shalit, and David Sontag.
\newblock Learning representations for counterfactual inference.
\newblock In {\em International conference on machine learning}, pages 3020--3029. PMLR, 2016.

\bibitem{fan2021causal}
Zhenjiang Fan, Kate~F Kernan, Panayiotis~V Benos, Gregory~F Cooper, Scott~W Canna, Joseph~A Carcillo, Soyeon Kim, and Hyun~Jung Park.
\newblock Causal inference using deep-learning variable selection identifies and incorporates direct and indirect causalities in complex biological systems.
\newblock {\em bioRxiv}, pages 2021--07, 2021.

\bibitem{louizos2017causal}
Christos Louizos, Uri Shalit, Joris~M Mooij, David Sontag, Richard Zemel, and Max Welling.
\newblock Causal effect inference with deep latent-variable models.
\newblock {\em Advances in neural information processing systems}, 30, 2017.

\bibitem{yoon2018ganite}
Jinsung Yoon, James Jordon, and Mihaela Van Der~Schaar.
\newblock Ganite: Estimation of individualized treatment effects using generative adversarial nets.
\newblock In {\em International conference on learning representations}, 2018.

\bibitem{lorch2022amortized}
Lars Lorch, Scott Sussex, Jonas Rothfuss, Andreas Krause, and Bernhard Sch{\"o}lkopf.
\newblock Amortized inference for causal structure learning.
\newblock {\em Advances in Neural Information Processing Systems}, 35:13104--13118, 2022.

\bibitem{schwab2018perfect}
Patrick Schwab, Lorenz Linhardt, and Walter Karlen.
\newblock Perfect match: A simple method for learning representations for counterfactual inference with neural networks.
\newblock {\em arXiv preprint arXiv:1810.00656}, 2018.

\bibitem{lim2018forecasting}
Bryan Lim.
\newblock Forecasting treatment responses over time using recurrent marginal structural networks.
\newblock {\em Advances in neural information processing systems}, 31, 2018.

\bibitem{sanchez2022causal}
Pedro Sanchez, Jeremy~P Voisey, Tian Xia, Hannah~I Watson, Alison~Q O’Neil, and Sotirios~A Tsaftaris.
\newblock Causal machine learning for healthcare and precision medicine.
\newblock {\em Royal Society Open Science}, 9(8):220638, 2022.

\bibitem{los_indicator}
Alex Bottle, Steven Middleton, Cor~J Kalkman, Edward~H Livingston, and Paul Aylin.
\newblock Global comparators project: international comparison of hospital outcomes using administrative data.
\newblock {\em Health services research}, 48(6pt1):2081--2100, 2013.

\bibitem{drugrepositioning}
Ted~T Ashburn and Karl~B Thor.
\newblock Drug repositioning: identifying and developing new uses for existing drugs.
\newblock {\em Nature reviews Drug discovery}, 3(8):673--683, 2004.

\bibitem{patientsimilarityanalysis}
Ahmed Allam, Matthias Dittberner, Anna Sintsova, Dominique Brodbeck, and Michael Krauthammer.
\newblock Patient similarity analysis with longitudinal health data.
\newblock {\em arXiv preprint arXiv:2005.06630}, 2020.

\bibitem{rl98}
Richard~S Sutton, Andrew~G Barto, et~al.
\newblock {\em Introduction to reinforcement learning}, volume 135.
\newblock MIT press Cambridge, 1998.

\bibitem{johnson2016mimic}
Alistair~EW Johnson, Tom~J Pollard, Lu~Shen, Li-wei~H Lehman, Mengling Feng, Mohammad Ghassemi, Benjamin Moody, Peter Szolovits, Leo Anthony~Celi, and Roger~G Mark.
\newblock Mimic-iii, a freely accessible critical care database.
\newblock {\em Scientific data}, 3(1):1--9, 2016.

\bibitem{johnson2016mimiciii}
A.~Johnson, T.~Pollard, and R.~Mark.
\newblock Mimic-iii clinical database (version 1.4).
\newblock \url{https://doi.org/10.13026/C2XW26}, 2016.

\bibitem{goldberger2000physiobank}
Ary~L Goldberger, Luis~AN Amaral, Leon Glass, Jeffrey~M Hausdorff, Plamen~Ch Ivanov, Roger~G Mark, Joseph~E Mietus, George~B Moody, Chung-Kang Peng, and H~Eugene Stanley.
\newblock Physiobank, physiotoolkit, and physionet: components of a new research resource for complex physiologic signals.
\newblock {\em circulation}, 101(23):e215--e220, 2000.

\bibitem{johnson2020mimiciv}
A.~Johnson, L.~Bulgarelli, T.~Pollard, S.~Horng, L.~A. Celi, and R.~Mark.
\newblock Mimic-iv (version 0.4).
\newblock \url{https://doi.org/10.13026/a3wn-hq05}, 2020.

\bibitem{attention2017}
Ashish Vaswani, Noam Shazeer, Niki Parmar, Jakob Uszkoreit, Llion Jones, Aidan~N Gomez, {\L}ukasz Kaiser, and Illia Polosukhin.
\newblock Attention is all you need.
\newblock {\em Advances in neural information processing systems}, 30, 2017.

\bibitem{intrinsiccomplexities}
Fernando Baquero and Cesar Nombela.
\newblock The microbiome as a human organ.
\newblock {\em Clinical Microbiology and Infection}, 18:2--4, 2012.

\bibitem{ICUprocedures}
Ghee~Chee Phua and Momen~M Wahidi.
\newblock Icu procedures of the critically ill.
\newblock {\em Respirology}, 14(8):1092--1097, 2009.

\bibitem{similarfeatures}
Jeffrey~H Silber, Paul~R Rosenbaum, Richard~N Ross, Justin~M Ludwig, Wei Wang, Bijan~A Niknam, Nabanita Mukherjee, Philip~A Saynisch, Orit Even-Shoshan, Rachel~R Kelz, et~al.
\newblock Template matching for auditing hospital cost and quality.
\newblock {\em Health services research}, 49(5):1446--1474, 2014.

\bibitem{careconsensus}
KCH Fearon, Olle Ljungqvist, M~Von~Meyenfeldt, A~Revhaug, CHC Dejong, K~Lassen, J~Nygren, J~Hausel, M~Soop, Jens Andersen, et~al.
\newblock Enhanced recovery after surgery: a consensus review of clinical care for patients undergoing colonic resection.
\newblock {\em Clinical nutrition}, 24(3):466--477, 2005.

\bibitem{postmanagement}
Henrik Kehlet, Roseanne~C Wilkinson, H~Barrie~J Fischer, Frederic Camu, Prospect~Working Group, et~al.
\newblock Prospect: evidence-based, procedure-specific postoperative pain management.
\newblock {\em Best practice \& research Clinical anaesthesiology}, 21(1):149--159, 2007.

\bibitem{healthcarereinforcement}
Chao Yu, Jiming Liu, Shamim Nemati, and Guosheng Yin.
\newblock Reinforcement learning in healthcare: A survey.
\newblock {\em ACM Computing Surveys (CSUR)}, 55(1):1--36, 2021.

\bibitem{sgld}
Max Welling and Yee~W Teh.
\newblock Bayesian learning via stochastic gradient langevin dynamics.
\newblock In {\em Proceedings of the 28th international conference on machine learning (ICML-11)}, pages 681--688, 2011.

\bibitem{AUC}
Jin Huang and Charles~X Ling.
\newblock Using auc and accuracy in evaluating learning algorithms.
\newblock {\em IEEE Transactions on knowledge and Data Engineering}, 17(3):299--310, 2005.

\bibitem{PRAUC}
Helen~R Sofaer, Jennifer~A Hoeting, and Catherine~S Jarnevich.
\newblock The area under the precision-recall curve as a performance metric for rare binary events.
\newblock {\em Methods in Ecology and Evolution}, 10(4):565--577, 2019.

\bibitem{F1scores}
Kanae Takahashi, Kouji Yamamoto, Aya Kuchiba, and Tatsuki Koyama.
\newblock Confidence interval for micro-averaged f 1 and macro-averaged f 1 scores.
\newblock {\em Applied Intelligence}, 52(5):4961--4972, 2022.

\bibitem{precisionrecall}
Cyril Goutte and Eric Gaussier.
\newblock A probabilistic interpretation of precision, recall and f-score, with implication for evaluation.
\newblock In {\em European conference on information retrieval}, pages 345--359. Springer, 2005.

\bibitem{Jaccard}
Luciano da~F Costa.
\newblock Further generalizations of the jaccard index.
\newblock {\em arXiv preprint arXiv:2110.09619}, 2021.

\bibitem{SSI-DDI}
Arnold~K Nyamabo, Hui Yu, and Jian-Yu Shi.
\newblock Ssi--ddi: substructure--substructure interactions for drug--drug interaction prediction.
\newblock {\em Briefings in Bioinformatics}, 22(6):bbab133, 2021.

\bibitem{hyperbolic}
Dmitri Krioukov, Fragkiskos Papadopoulos, Maksim Kitsak, Amin Vahdat, and Mari{\'a}n Bogun{\'a}.
\newblock Hyperbolic geometry of complex networks.
\newblock {\em Physical Review E}, 82(3):036106, 2010.

\bibitem{hyperbolicmapping}
Mari{\'a}n Bogun{\'a}, Fragkiskos Papadopoulos, and Dmitri Krioukov.
\newblock Sustaining the internet with hyperbolic mapping.
\newblock {\em Nature communications}, 1(1):62, 2010.

\bibitem{dimensionality}
Ella Bingham and Heikki Mannila.
\newblock Random projection in dimensionality reduction: applications to image and text data.
\newblock In {\em Proceedings of the seventh ACM SIGKDD international conference on Knowledge discovery and data mining}, pages 245--250, 2001.

\bibitem{nonlinear}
John~Aldo Lee, Amaury Lendasse, and Michel Verleysen.
\newblock Nonlinear projection with curvilinear distances: Isomap versus curvilinear distance analysis.
\newblock {\em Neurocomputing}, 57:49--76, 2004.

\bibitem{hyperbolicspace}
Alessandro Muscoloni, Josephine~Maria Thomas, Sara Ciucci, Ginestra Bianconi, and Carlo~Vittorio Cannistraci.
\newblock Machine learning meets complex networks via coalescent embedding in the hyperbolic space.
\newblock {\em Nature communications}, 8(1):1615, 2017.

\bibitem{umap}
Leland McInnes, John Healy, and James Melville.
\newblock Umap: Uniform manifold approximation and projection for dimension reduction.
\newblock {\em arXiv preprint arXiv:1802.03426}, 2018.

\bibitem{Poincaré}
Maximillian Nickel and Douwe Kiela.
\newblock Learning continuous hierarchies in the lorentz model of hyperbolic geometry.
\newblock In {\em International conference on machine learning}, pages 3779--3788. PMLR, 2018.

\bibitem{poincareclustered}
Anna Klimovskaia, David Lopez-Paz, L{\'e}on Bottou, and Maximilian Nickel.
\newblock Poincar{\'e} maps for analyzing complex hierarchies in single-cell data.
\newblock {\em Nature communications}, 11(1):2966, 2020.

\bibitem{kmeans1982}
Stuart Lloyd.
\newblock Least squares quantization in pcm.
\newblock {\em IEEE transactions on information theory}, 28(2):129--137, 1982.

\bibitem{tf_idf_ref}
Claude Sammut and Geoffrey~I. Webb, editors.
\newblock {\em TF--IDF}, pages 986--987.
\newblock Springer US, Boston, MA, 2010.

\bibitem{pyhealth2023yang}
Chaoqi Yang, Zhenbang Wu, Patrick Jiang, Zhen Lin, Junyi Gao, Benjamin Danek, and Jimeng Sun.
\newblock {PyHealth}: A deep learning toolkit for healthcare predictive modeling.
\newblock In {\em Proceedings of the 27th ACM SIGKDD International Conference on Knowledge Discovery and Data Mining (KDD) 2023}, 2023.

\bibitem{zollner2007opioids}
C~Z{\"o}llner and C~Stein.
\newblock Opioids.
\newblock {\em Analgesia}, pages 31--63, 2007.

\bibitem{weersma2020interaction}
Rinse~K Weersma, Alexandra Zhernakova, and Jingyuan Fu.
\newblock Interaction between drugs and the gut microbiome.
\newblock {\em Gut}, 69(8):1510--1519, 2020.

\bibitem{al2022hypertension}
Akram Al-Makki, Donald DiPette, Paul~K Whelton, M~Hassan Murad, Reem~A Mustafa, Shrish Acharya, Hind~Mamoun Beheiry, Beatriz Champagne, Kenneth Connell, Marie~Therese Cooney, et~al.
\newblock Hypertension pharmacological treatment in adults: a world health organization guideline executive summary.
\newblock {\em Hypertension}, 79(1):293--301, 2022.

\bibitem{an2023comprehensive}
Qi~An, Saifur Rahman, Jingwen Zhou, and James~Jin Kang.
\newblock A comprehensive review on machine learning in healthcare industry: Classification, restrictions, opportunities and challenges.
\newblock {\em Sensors}, 23(9):4178, 2023.

\bibitem{miltzow2022classifying}
Tillmann Miltzow and Reinier~F Schmiermann.
\newblock On classifying continuous constraint satisfaction problems.
\newblock In {\em 2021 IEEE 62nd Annual Symposium on Foundations of Computer Science (FOCS)}, pages 781--791. IEEE, 2022.

\bibitem{zhu2021overview}
Jie Zhu, Fengge Wu, and Junsuo Zhao.
\newblock An overview of the action space for deep reinforcement learning.
\newblock In {\em Proceedings of the 2021 4th International Conference on Algorithms, Computing and Artificial Intelligence}, pages 1--10, 2021.

\bibitem{walker2022program}
Simon Walker, Aimee Fox, James Altunkaya, Tim Colbourn, Mike Drummond, Susan Griffin, Nils Gutacker, Paul Revill, and Mark Sculpher.
\newblock Program evaluation of population-and system-level policies: evidence for decision making.
\newblock {\em Medical Decision Making}, 42(1):17--27, 2022.

\bibitem{nanayakkara2022unifying}
Thesath Nanayakkara, Gilles Clermont, Christopher~James Langmead, and David Swigon.
\newblock Unifying cardiovascular modelling with deep reinforcement learning for uncertainty aware control of sepsis treatment.
\newblock {\em PLOS Digital Health}, 1(2):e0000012, 2022.

\bibitem{raghu2017continuous}
Aniruddh Raghu, Matthieu Komorowski, Leo~Anthony Celi, Peter Szolovits, and Marzyeh Ghassemi.
\newblock Continuous state-space models for optimal sepsis treatment: a deep reinforcement learning approach.
\newblock In {\em Machine Learning for Healthcare Conference}, pages 147--163. PMLR, 2017.

\bibitem{scali2015primary}
Elena~P Scali, Tracy~M Chandler, Eric~J Heffernan, Joseph Coyle, Alison~C Harris, and Silvia~D Chang.
\newblock Primary retroperitoneal masses: what is the differential diagnosis?
\newblock {\em Abdominal imaging}, 40:1887--1903, 2015.

\bibitem{lecun2015deep}
Yann LeCun, Yoshua Bengio, and Geoffrey Hinton.
\newblock Deep learning.
\newblock {\em nature}, 521(7553):436--444, 2015.

\bibitem{minervini2023adaptive}
Pasquale Minervini, Luca Franceschi, and Mathias Niepert.
\newblock Adaptive perturbation-based gradient estimation for discrete latent variable models.
\newblock In {\em Proceedings of the AAAI Conference on Artificial Intelligence}, volume~37, pages 9200--9208, 2023.

\bibitem{yu2019gradient}
Da~Yu, Huishuai Zhang, Wei Chen, Tie-Yan Liu, and Jian Yin.
\newblock Gradient perturbation is underrated for differentially private convex optimization.
\newblock {\em arXiv preprint arXiv:1911.11363}, 2019.

\bibitem{agarwal2021towards}
Sushant Agarwal, Shahin Jabbari, Chirag Agarwal, Sohini Upadhyay, Steven Wu, and Himabindu Lakkaraju.
\newblock Towards the unification and robustness of perturbation and gradient based explanations.
\newblock In {\em International Conference on Machine Learning}, pages 110--119. PMLR, 2021.

\bibitem{hazan2016perturbations}
Tamir Hazan, George Papandreou, and Daniel Tarlow.
\newblock {\em Perturbations, optimization, and statistics}.
\newblock MIT Press, 2016.

\end{thebibliography}

\end{document}